\documentclass[fleqn,usenatbib]{mnras}

\usepackage[T1]{fontenc}
\usepackage{ae,aecompl}
\usepackage{amsmath}
\usepackage{natbib}
\usepackage{graphicx}
\usepackage{xcolor}
\usepackage{amssymb}
\usepackage{newtxtext,newtxmath}

\graphicspath{{./}{figures/}}

\title[Old Inner Galaxy Stars in FIRE-2 Simulations]{The Dynamics of Old Inner Galaxy Stars in Milky Way-mass Galaxies Using FIRE-2 Simulations}

\author[Tasso et al.]{
Emma Tasso,$^{1,2}$
Madeline Lucey,$^{1}$
Robyn Sanderson,$^{1}$
Aritra Kundu,$^{1}$
and Lina Necib$^{3}$
\\
$^{1}$Department of Physics \& Astronomy, University of Pennsylvania, 209 S 33rd Street, Philadelphia, PA 19104, USA\\
$^{2}$Graduate Center, City University of New York, 365 5th Avenue, New York, NY 10016, USA\\
$^{3}$Department of Physics and Kavli Institute for Astrophysics and Space Research, Massachusetts Institute of Technology,\\ 77 Massachusetts Avenue, Cambridge, MA 02139, USA
}

\date{Accepted XXX. Received YYY; in original form ZZZ}

\pubyear{\the\year{}}

\begin{document}

\label{firstpage}
\pagerange{\pageref{firstpage}--\pageref{lastpage}}
\maketitle






\begin{abstract}

Understanding how galaxies like the Milky Way assembled over cosmic time remains a central question in astrophysics. Understanding the processes that shaped their formation and evolution is greatly enhanced by the joint use of observational data and high-resolution cosmological simulations. Old stars in the inner regions of galaxies serve as powerful tracers of early dynamical events, having formed during the initial stages of galaxy assembly and retaining the kinematic imprints of those formative periods. We investigate the kinematic properties of old (age $>10$ Gyr) inner galaxy ($r_\mathrm{GC}$
 $<5$ kpc) stars in thirteen Milky Way-mass galaxies from the FIRE-2 cosmological zoom-in simulations, focusing on their origin, orbital structure, and kinematic alignment with the disk. Our analysis reveals that old stars in the inner galaxy are more likely to have been formed in their host galaxy, although accretion is seen most prominently during the earliest stages of galaxy formation. Many of these accreted stars tend to occupy kinematically hot orbits compared to their counterparts formed in the host galaxy, although some stars formed in-galaxy also retain kinematically hot orbits. Disk-like dynamics are present throughout all age bins, and are most prominent as age decreases. Although some old stellar populations retain disk-like structure, the prominence of this rotational component varies significantly across galaxies and between star populations. These results emphasize the diversity of early galaxy assembly histories and suggest that coherent angular momentum in accreted material can leave detectable kinematic signatures in present-day stellar halos.

\end{abstract}

\begin{keywords}
stars: Population III -- galaxies: bulges -- galaxies:
kinematics and dynamics
\end{keywords}

\section{Introduction} \label{sec:intro}

How did galaxies like the Milky Way assemble over cosmic time? This fundamental question lies at the heart of modern astrophysics and motivates the present study. Our location within the Milky Way offers a unique vantage point, allowing us to observe its structure in detail, compare against computational models, and test predictions in our own galactic backyard. 

Understanding the formation and evolution of stellar halos and disks in Milky Way-mass galaxies is central to reconstructing the assembly history of galaxies. In particular, old (age $>10$ Gyr) inner galaxy ($r_\mathrm{GC}<5$ kpc) stars serve as valuable tracers of early dynamical events and can retain signatures of both internal galaxy formation and past accretion \citep{Elbadry2018}. The stellar halo, composed largely of ancient metal-poor stars, preserves critical clues about early galaxy assembly and high-redshift star formation \citep{Freeman2002, Beers2005, Frebel2015, Bullock2001, Bullock2005, Salvadori2008, Helmi2008, Kirby2008b,Bovill2011, Brown2014, Beniamini2018, Magg2017, Tumlinson2010, El-Badry2018a}. Studies using non-cosmological N-body simulations, when compared to observations of the Milky Way halo, have been shown to realistically reproduce its basic structural and kinematic properties \citep{Robertson2005, Font2006a}. \citet{Rix2022} used Gaia DR3 to identify a centrally concentrated, metal-poor population of old stars in the inner few kiloparsecs of the Milky Way, many of which show high eccentricities and weak net rotation. This supports the concept of an ancient proto‑Galaxy component that our FIRE‑2 simulation-based analysis aims to model and interpret \citep{Rix2022}.

The orbital properties of old inner galaxy stars, such as their eccentricities, circularities, and vertical excursions, can reveal whether they formed within the main galaxy or were brought in through mergers. Stars accreted from satellite galaxies are typically expected to exhibit more dispersion‑dominated spheroidal kinematics, while in‑situ stars may retain coherent disk‑like motion if formed in rotationally supported environments \citep{Elbadry2018}. However, recent simulation studies suggest that this distinction is not always clear. For example, studies have shown that some accreted stars can exhibit disk-like orbits \citep{Sanderson2020, Yu2023}. Moreover, \citet{Horta2024} show that proto‑Galactic components in FIRE‑2 analogs can emerge from multiple similarly massive progenitors and sometimes retain weak net rotation aligned with the present‑day disk, signaling a more nuanced connection between hierarchical assembly and disk formation. 

It is generally agreed that outer halo stars come from largely accreted origins \citep{Bullock2005, Helmi2008, Bullock2001, Bell2008, Monachesi2019}. The inner stellar halo is more nuanced, with contributions from both accreted \citep{Helmi2018, Mackereth2018} and in-situ stars \citep{Carollo2007, Cooper2015}. The dispersion-dominated kinematics of this in-situ halo population likely arise from dynamical heating after forming in the galactic disk \citep{Zolotov2009, Purcell2010}, or from feedback-driven processes that altered their orbits \citep{Elbadry2018}. The idea that stellar halos are primarily assembled through accretion is supported by both theoretical predictions from standard dark energy/cold dark matter ($\Lambda$CDM) cosmology \citep{White1991, Lucey2025, Bullock2005, Cooper2010} and observational evidence, such as tidal streams detected around nearby galaxies \citep{Belokurov2006, McConnachie2009, Martinez-Delgado2010}.

Hydrodynamical simulations consistently produce distinct in-situ halo components as a consequence of galaxy formation in the $\Lambda$CDM framework \citep{Abadi2003, Brook2004, Zolotov2009, Font2011, Tissera2013, Pillepich2015, Lucey2025}. In this model, most stars in a Milky Way–like galaxy are expected to form from gas that cools within the galaxy’s own dark matter potential well \citep{White1978, White1991}. Most in-situ stars belong to the cold, rotating Galactic disk, although the proto-Milky Way likely experienced significant satellite interactions and accretion events \citep{Bett2012}, or even full disk disruption and subsequent regrowth prior to the formation of most present-day disk stars \citep{Sales2012, Aumer2013a, Aumer2013b}.

Several prior studies using FIRE-2 have investigated early galaxy assembly from complementary angles. \citet{Horta2024} study the proto-galaxy populations of the same 13 Milky Way-mass FIRE-2 galaxies, identifying the main branch and building block systems that coalesce before the galaxy becomes dynamically dominant, defined by a 3:1 stellar mass ratio with the next most massive halo (i.e. $t_{\rm MR3:1}$\footnote{This is the redshift at which the most massive halo in the main branch reaches a stellar mass that is three times that of the second most massive halo.}). They find that proto-Milky Way populations are composed of one or two dominant LMC-mass systems and three to five lower-mass building blocks, and that these populations show weak but systematic prograde rotation relative to the present-day disc. \citet{Yu2023} similarly use FIRE-2 to study the kinematic history of in-situ stars, classifying them into structural components (spheroid, thick disc, and thin disc) using orbital circularity, and tracing how the galaxy transitions from a bursty, dispersion-dominated phase to a settled rotating disc. While these studies provide important context for early galaxy formation, \citet{Horta2024} focus primarily on distinguishing the massive progenitor systems that built the proto-galaxy rather than comparing the full accreted-versus-in-situ population at fixed age and radius, and \citet{Yu2023} restrict their analysis to in-situ stars and disk formation rather than examining accreted populations and their present-day kinematics. Our work provides the theoretical complement to observational studies of the dynamics of old metal-poor stars in the inner galaxy \citep{Lucey2021, Ardern-Arentsen2024}. 

In this work, we build directly on these studies by examining the present-day kinematic properties of all old (age $> 10$ Gyr) inner galaxy ($r_{\rm GC} < 5$ kpc) stars, both accreted and in-situ, across all thirteen FIRE-2 Milky Way-mass galaxies, using a symmetry-based diskiness metric to quantify disk-like dynamics at the population level. This allows us to ask not just how the proto-galaxy was assembled, or how the disk settled, but how both of those histories are imprinted in the orbital structure of the oldest stars observable today in the inner galaxy.

\section{FIRE-2 Simulations and Milky Way-like Galaxies} \label{sec:fire}

This work utilizes the FIRE-2 cosmological zoom-in simulations. We focus on the Milky Way-mass galaxies in  FIRE-2, which serve as realistic analogs to the Milky Way and enable us to make testable predictions. Specifically, we examine seven isolated Milky Way-mass galaxies within the \textit{Latte} suite: m12i, m12b, m12c, m12f, m12m, m12r, and m12w \citep{Wetzel2023,Wetzel2025}. We also examine three Local Group-like pairs within the ELVIS suite: Romeo \& Juliet, Romulus \& Remus, and Thelma \& Louise \citep{Garrison-Kimmel2019}.

All simulations use the FIRE-2 galaxy formation model \citep{Hopkins2018b} and are run with the GIZMO gravity and hydrodynamics code \citep{Hopkins2015, GizmoAnalysis}, employing its meshless finite-mass (MFM) solver. Each simulation adopts a flat $\Lambda$CDM cosmology with parameters consistent with those reported by \citet{Planck2014}.

The \textit{Latte} suite, excluding m12w, uses $\Omega_{\mathrm{m}} = 0.272$, $\Omega_{\mathrm{b}} = 0.0455$, $\sigma_8 = 0.807$, $n_{\mathrm{s}} = 0.961$, and $h = 0.702$. Of the ELVIS suite of galaxy pairs, Thelma \& Louise and Romulus \& Remus both utilize the same cosmology as the original ELVIS dark-matter-only (DMO) suite: $\Omega_{\mathrm{m}} = 0.266$, $\Omega_{\mathrm{b}} = 0.0449$, $\sigma_8 = 0.801$, $n_{\mathrm{s}} = 0.963$, and $h = 0.71$. The third ELVIS galaxy pair, Romeo \& Juliet, along with one \textit{Latte} galaxy, m12w, both use updated parameters from \citet{Planck2020}: $\Omega_{\mathrm{m}} = 0.31$, $\Omega_{\mathrm{b}} = 0.048$, $\sigma_8 = 0.82$, $n_{\mathrm{s}} = 0.97$, and $h = 0.68$ .

The feedback mechanisms implemented in FIRE-2 significantly influence the mass distributions of these Milky Way analogs \citep{Pontzen2014,Lazar2020}. The FIRE-2 simulations incorporate stellar feedback from processes such as stellar winds, radiation pressure from young stars, Type II and Ia supernovae, photoelectric heating, and photoionization heating, all of which help regulate star formation. The gas density threshold for star formation in FIRE-2 is \( n_{\mathrm{SF}} > 1000\ \mathrm{cm}^{-3} \). Feedback event rates, luminosities, energies, mass-loss rates, and related quantities are derived directly from stellar evolution model outputs \citep{Leitherer1999}. 

All isolated and paired galaxies have dark matter halo masses at $z = 0$ of $M_{200} = 1$--$2.1 \times 10^{12}~M_\odot$ \citep{Sanderson2018}. The Latte suite galaxies have an initial stellar particle mass of $7070~M_\odot$, whereas the ELVIS suite galaxies are simulated at higher resolution with stellar particle masses of $3500~M_\odot$. The gravitational softening lengths are approximately $4~\mathrm{pc}$ for star particles and $40~\mathrm{pc}$ for dark matter particles.

These simulated galaxies reproduce the observed stellar mass--halo mass relation across cosmic time \citep{Hopkins2018}. They are broadly consistent with several key properties of the Milky Way, including the stellar halo mass fraction \citep{Sanderson2018}, the presence of a metal-rich in-situ halo component \citep{Bonaca2017}, the radial and vertical structure of the stellar disk \citep{Ma2017,Sanderson2020,Bellardini2021,McCluskey2024}, and the observed satellite populations around the Milky Way and M31 \citep{Wetzel2016,Samuel2020,Garrison-Kimmel2019b,Panithanpaisal2021,Cunningham2021}. However, some discrepancies remain: for instance, \citet{Shipp2023} find that while the masses and number of stellar streams are consistent with observations, their orbital properties are not. Additionally, FIRE-2 disks form later and are therefore younger than current estimates of the age of the Milky Way disk \citep{McCluskey2024}.

To define the center of each galaxy, we follow the method described in \citet{Santistevan2020}, where the center-of-mass position is calculated using star particles. Specifically, we identify the position that encloses 90\% of the stellar mass within $1.5\,R_{\mathrm{star,90}}$ and iteratively recenter both the galaxy and the halo such that the galaxy's center-of-mass position coincides with that of the host halo. This ensures a consistent definition of the galactic center across snapshots and simulations. Since the formation distance is measured relative to the center of the host galaxy, defining this center based on the stellar mass distribution ensures a physically meaningful and consistent reference point throughout the simulations.

To distinguish between accreted and in-situ star particles, we use the ROCKSTAR halo catalogs and merger trees \citep{Behroozi2013a,Behroozi2013b}, along with a substructure tracking and identification method introduced in \citet{Panithanpaisal2021} and updated in \citet{Kundu2025}. We track each bound subhalo with non-zero stellar mass and its associated star particles that falls into the virial radius ($R_{\rm 200m}$) of the main progenitor (most massive progenitor) of the present-day host galaxy. We begin tracking stellar histories from snapshot~70, which corresponds to a cosmic time of $t = 1.26\,\mathrm{Gyr}$. This choice reflects a balance between probing early galaxy assembly and maintaining a feasible computational cost. Only a small fraction of stars in our old inner 
galaxy sample formed prior to this time, with an average of $3.8\%$ across galaxies. Just three systems exceed $5\%$, and the maximum fraction is $7.3\%$ in Romulus, indicating that our choice of starting snapshot does not significantly bias our results. The star particles belonging to these subhalos are defined as accreted star particles, while the stars that formed in the main halo before we start to track are considered to be part of the in-situ particles. Unlike \citet{Panithanpaisal2021}, who further classify the accreted substructures -- with stellar mass between $10^6 \, M_\odot \lesssim M_\star \lesssim 10^9 \, M_\odot$ -- into satellites, stellar streams, or phase-mixed debris, we keep all the accreted objects. This leads us to include very massive objects ($M_\star \sim 5 \times 10^9 \, M_\odot$), which are morphologically complex  \citep{Kundu2025} and are missing in \citet{Panithanpaisal2021}.

\section{Star Selection} \label{sec:star}

\begin{figure*}
    \centering
    \includegraphics[width=\linewidth]{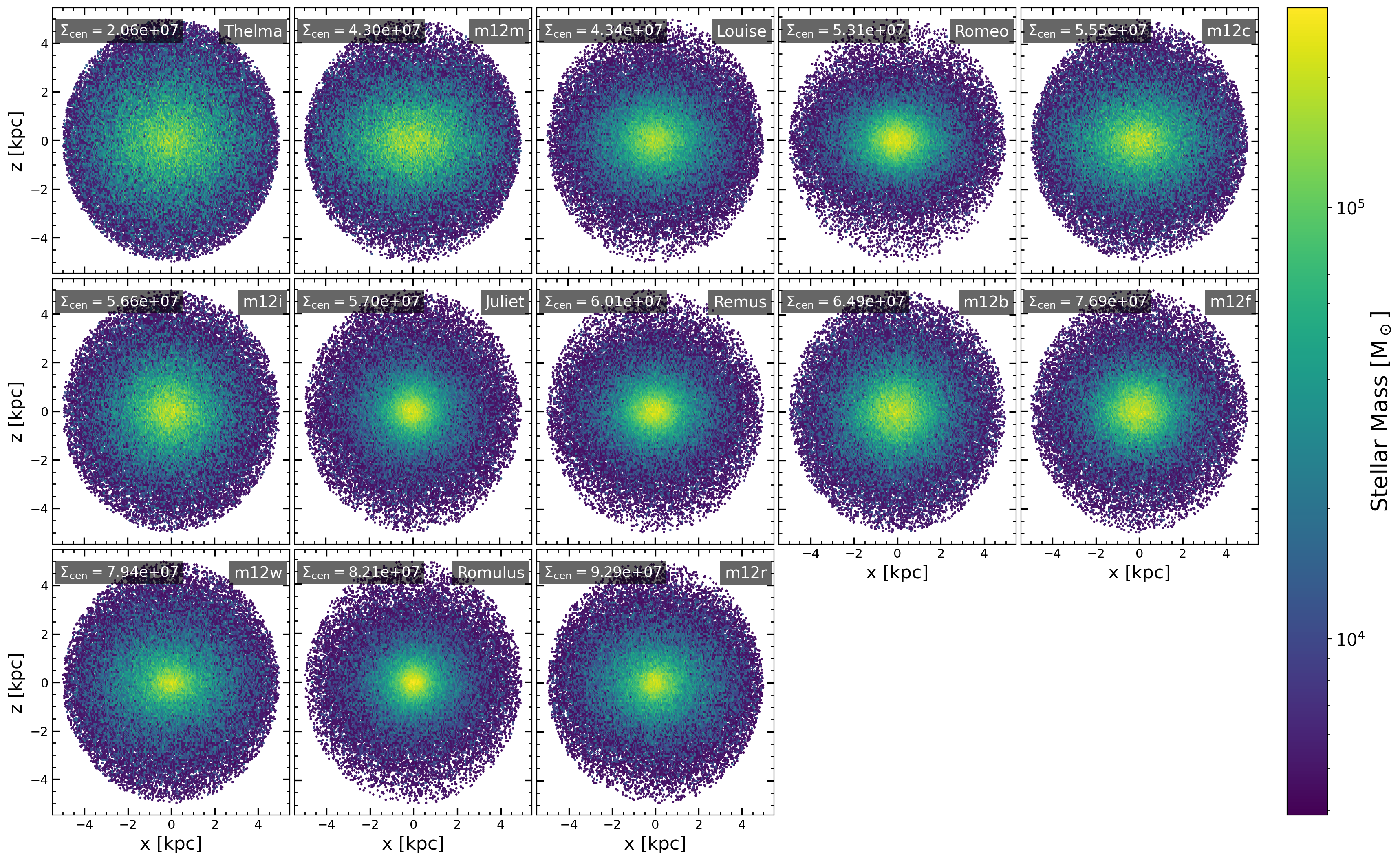}
    \caption{
    Stellar surface density maps in the $x$--$z$ plane for old (age $>10$ Gyr) inner galaxy ($r_{\mathrm{GC}} < 5$ kpc) stars in each FIRE--2 simulation. Each galaxy is oriented such that its disk lies in the $x$--$y$ plane, and the origin marks the center of mass of the galaxy. The quantity $\Sigma_{\mathrm{cen}}$ [M$_\odot$ kpc$^{-2}$] shown in each panel is the mean surface density of the ten highest-density hexbins, providing a robust estimate of the central stellar concentration. Galaxies are ordered from lowest to highest $\Sigma_{\mathrm{cen}}$, such that Thelma is the most diffuse and m12r the most centrally concentrated. Color indicates stellar mass per bin.}
    \label{fig:density}
\end{figure*}

We focus on old (age $>$10 Gyr) star particles located within the inner galaxy, defined as a spherical region of radius 5~kpc centered on each galaxy's center of mass. To ensure computational feasibility while maintaining robust statistics, we randomly selected 100,000 stars from the old stellar population within this volume for each simulation. The breakdown of selected accreted and non-accreted particles is shown in Table~\ref{tab:combined_table}, and the spatial distribution is visualized in Figure~\ref{fig:density}.

Figure~\ref{fig:density} provides an overview of the selected star sample for each galaxy, ordered by mean central stellar surface density ($\Sigma_{\mathrm{cen}}$). This ordering highlights structural differences across the sample: some galaxies exhibit compact, centrally concentrated distributions (e.g., m12r), while others are more diffuse (e.g., Thelma). We use this ordering in some subsequent figures. The corresponding values of the central surface density are reported in Table~\ref{tab:combined_table}. 

\section{Orbit Calculations} \label{sec:orbit}

We integrated stellar orbits using the AGAMA \citep{Vasiliev2021} framework. For each FIRE-2 galaxy, we construct a static multicomponent potential from the $z=0$ snapshot. Rather than fitting separate analytic disk, bulge, and halo profiles, we approximate the simulated mass distribution using basis-function expansions. The dark matter halo and hot gas are represented with a low-order spherical-harmonic/multipole expansion, while the baryonic material, including the central stellar component, flattened stellar disk, and cold gas, is represented with an azimuthal harmonic expansion in cylindrical coordinates, following \citet{Arora2022, Arora2024b}. These expansion components are summed to form the total potential. For paired systems, independent potentials were built for each host. We then integrate the orbits for 1 Gyr in this fixed $z=0$ potential using 1000 timesteps, rather than evolving the potential across multiple simulation snapshots. This duration is sufficient to sample multiple orbital periods of old inner-galaxy stars.

Because some FIRE-2 galaxies may experience recent massive mergers, these orbit integrations should not be interpreted as exact reconstructions of each star particle's time-dependent past trajectory. Instead, our goal is to perform an observationally motivated dynamical experiment: given the present-day phase-space coordinates of stars and a present-day galaxy potential, we ask what orbital properties would be inferred from integrating those stars in a fixed potential.

\begin{table*}
\centering
\small
\setlength{\tabcolsep}{4pt}
\begin{tabular}{lccccccc}
\hline
\textbf{Simulation} & \(\mathbf{M_{200\mathrm{c}}}\) [$10^{10}\ M_\odot$] & \(\mathbf{R_{200\mathrm{c}}}\) [kpc] & $\mathbf{t_{{MR}_{3:1}}}$ [Gyr]
 & \textbf{Accretion Ratio [\%]} & {\boldmath$\Sigma_{\mathrm{cen}}$} [$10^{7}\,M_\odot\,\mathrm{kpc}^{-2}$] & $\boldsymbol{\langle \epsilon \rangle}$ & \textbf{Diskiness Fraction [\%]} \\
\hline
Juliet    & 72.4 & 189 & 12.55 & 1.86 & 5.696 & 0.017 & 3.4 \\
m12i      & 78.3 & 190 & 11.80  & 4.88 & 5.664 & 0.032 & 5.9 \\
Louise    & 71.0 & 182 & 11.90 & 2.65 & 4.343 & 0.046 & 8.7 \\
Thelma    & 94.1 & 200 & 12.40 & 55.18 & 2.057 & 0.027 & 9.3 \\
m12b      & 94.2 & 202 & 9.76 & 10.16 & 6.494 & 0.049 & 11.6 \\
Romulus   & 134  & 225 & 10.00 & 22.67 & 8.208 & 0.053 & 12.0 \\
Remus     & 85.9 & 194 & 7.90 & 7.65 & 6.008 & 0.058 & 12.1 \\
m12f      & 108  & 211 & 8.05 & 1.69 & 7.687 & 0.080 & 16.9 \\
m12c      & 91.2 & 200 & 9.05 & 50.69 & 5.554 & 0.081 & 17.3 \\
m12r      & 78.5 & 194 & 8.05 & 4.15 & 9.287 & 0.090 & 17.5 \\
m12w      & 71.8 & 188 & 11.60 & 7.52 & 7.940 & 0.144 & 29.0 \\
m12m      & 97.8 & 204 & 9.94 & 35.09 & 4.302 & 0.167 & 35.1 \\
Romeo     & 93.9 & 206 & 12.81 & 0.15 & 5.310 & 0.190 & 37.6 \\
\hline
\end{tabular}

\caption{Summary of host galaxy properties, accretion ratios, central stellar surface densities, and total diskiness fractions for each FIRE-2 galaxy at \(z=0\). \(M_{200\mathrm{c}}\) and \(R_{200\mathrm{c}}\) are the total mass and spherical radius within which the average density is 200 times the critical density of the universe and $t_{\mathrm{MR}_{3:1}}$ is the approximate times in the simulation in which we define the proto-Milky Way to form. The accretion ratio is computed as the ratio of accreted stars to stars formed in the host galaxy for old (age \(>10\) Gyr) inner galaxy ($r_\mathrm{GC}$ $<5$ kpc) stars across all times, and is reported here as a percentage. Central surface densities ($\Sigma_{\mathrm{cen}}$) are computed as the mean stellar surface density of the ten highest-density bins in the \(x\)–\(z\) projection. The mean circularity, $\langle \epsilon \rangle$, is computed for the old inner-galaxy stellar population. The diskiness fraction is computed using the difference method described in Figure~\ref{fig:flip}, and is reported here as a percentage. The accretion ratio, surface density, mean circularity, and diskiness fraction are calculated from the old inner galaxy selection only.}
\label{tab:combined_table}
\end{table*}

\begin{figure*}
    \centering
    \includegraphics[width=\linewidth]{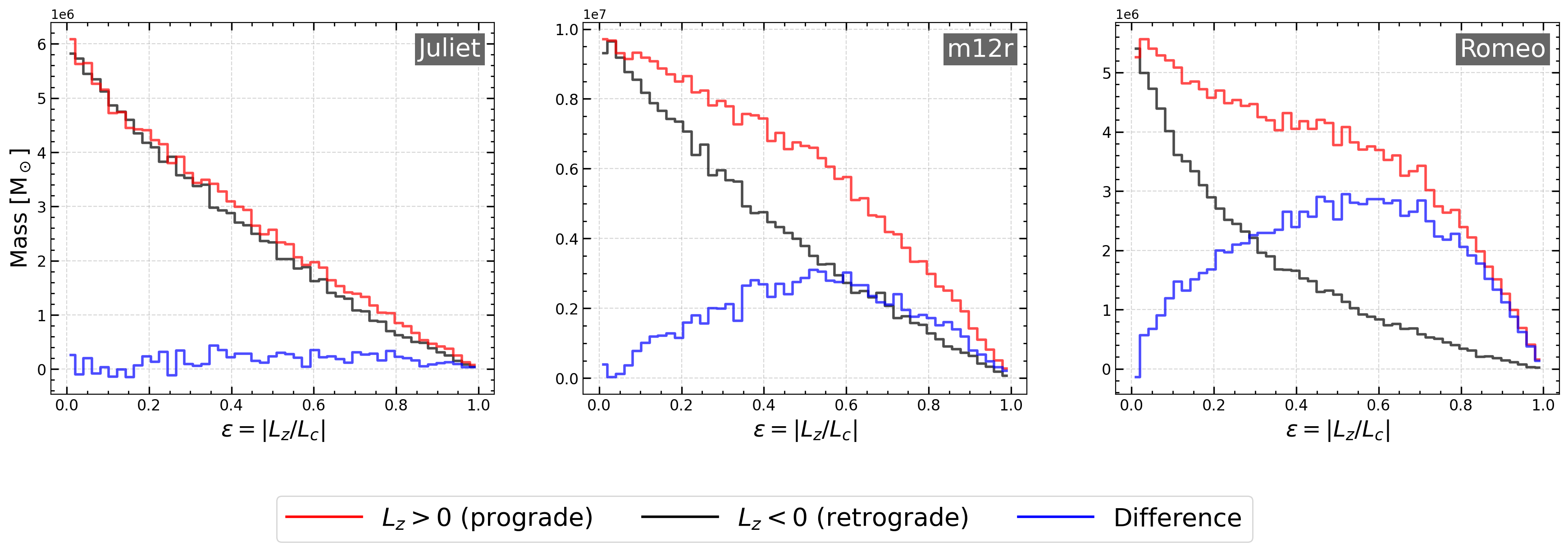}
\caption{Mass weighted orbital circularity distributions for old inner galaxy stars in the Juliet, m12r, and Romeo simulations, following the notation used in \citet{Yu2023}. The red and black lines represent mass-weighted histograms of $\epsilon = L_z/L_c$ for stars on prograde and retrograde orbits, respectively. The blue line shows the mass difference between the two, highlighting the net rotating component associated with a disk. Juliet shows a low diskiness fraction with a relatively flat difference distribution, m12r exhibits a modest central peak, and Romeo has a strongly peaked blue distribution consistent with a high diskiness fraction.}
    \label{fig:flip}
\end{figure*}

To assess the orbits of old inner galaxy stars, we analyze the orbital circularity histograms for each simulation, with representative examples shown in Figure~\ref{fig:flip}. 
Orbital circularity is defined as:
\begin{equation}
\epsilon = \frac{L_z}{L_c},
\end{equation}
where $L_z$ is the specific angular momentum in the $z$ direction and $L_c$ is the angular momentum of a circular orbit at the same energy. The $\hat{z}$ direction is defined by the total stellar angular momentum within 10 kpc of each galaxy's center at the snapshot of interest, here $z=0$. We confirm that the net angular momentum of stars within 5 kpc is highly aligned with the $z$ direction in all simulations, with $|L_z|/|\vec{L}_{\rm inner}| > 0.99$. A value of $\epsilon=1$ corresponds to a circular orbit in the disk plane, while $\epsilon<0$ indicates counter-rotation. For the representative systems shown in Figure~\ref{fig:flip}, we compare the stellar mass in prograde orbits ($\epsilon>0$; red) to the stellar mass in retrograde orbits ($\epsilon<0$; black), with their difference shown in blue.

The underlying idea is that, while individual non-disk stars may have non-zero angular momentum, their orbits are more randomly oriented, so the net rotation of the dipersion dominated stellar population should be close to zero. As a result, the distribution of circularity ($\epsilon$) values is expected to be symmetric around zero if it is composed solely of non-disk stars. Any excess toward positive $\epsilon$ values indicates a population of stars with coherent prograde rotation, likely reflecting the presence of a disk-like component. This metric does not classify individual stars as disk or non-disk, but instead characterizes the overall population by identifying the fraction of stars that exhibit disk-like kinematics.

In Figure~\ref{fig:flip}, we highlight three simulations: Juliet, which shows a nearly symmetric distribution; Romeo, which exhibits a strong positive excess; and m12r, which represents an intermediate case. To quantify the asymmetry in these distributions for each simulation, we compute the fraction of stars with disk-like dynamics as:
\begin{equation}
\text{Diskiness Fraction} = \frac{M(\epsilon > 0) - M(\epsilon < 0)}{M_{\text{total}}}.
\end{equation}

A FIRE-2 galaxy with a high diskiness fraction contains a larger proportion of stars whose orbits resemble those of a rotationally supported disk. A low diskiness fraction indicates that most stars follow more random, halo-like motions rather than organized disk dynamics. This metric does not directly classify stellar populations as belonging to a disk, but instead describes how disk-like their overall kinematics are.
These values are listed in Table~\ref{tab:combined_table}. 

We emphasize that this metric does not classify individual stars with $\epsilon>0$ as disk stars. Instead, it measures the population-level excess of prograde over retrograde stellar mass. Thus, our use of ``diskiness'' refers to the degree of net coherent prograde rotation in the population, rather than the fraction of stars on cold, thin-disk-like orbits. Because the diskiness fraction depends only on the prograde-retrograde mass imbalance, we also report the mean circularity, $\langle \epsilon \rangle$, as a complementary measure of the typical circularity values in each galaxy. Because all circularities are measured at $z=0$, the prograde excess of old stars should be interpreted as a present-day dynamical signature. It does not by itself determine whether these stars formed on rotating orbits or were later torqued into alignment with the disk.

\begin{figure}
    \centering
    \includegraphics[width=\linewidth]{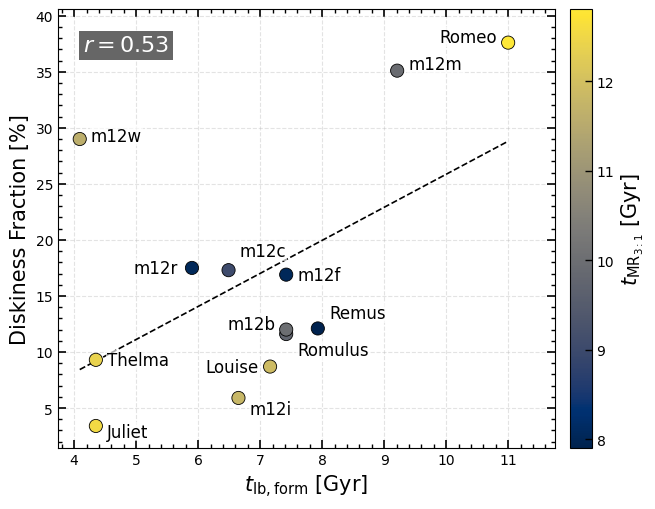}
    \caption{
    Diskiness fraction compared to the disk-settling lookback time, $t_{\rm lb,form}$, from \citet{McCluskey2024}. Points are colored by $t_{\rm MR_{3:1}}$, the approximate lookback time at which we define the proto-Milky Way to form, with values also reported in Table~\ref{tab:combined_table}. The dashed line shows a linear fit, and the Pearson correlation coefficient, $r$, is shown in the upper left; positive values of $r$ indicate that larger disk-settling lookback times are associated with larger diskiness fractions. Larger values of both $t_{\rm lb,form}$ and $t_{\rm MR_{3:1}}$ correspond to earlier times in the galaxy's evolution.}
    
    \label{fig:diskiness_settling}
\end{figure}

To explore whether the present-day prograde excess of old inner-galaxy stars is connected to the timing of disk settling, we compare our diskiness fractions to the disk-settling lookback time, $t_{\rm lb,form}$, from \citet{McCluskey2024} in Figure~\ref{fig:diskiness_settling}. We color the points by $t_{\rm MR_{3:1}}$, the approximate lookback time at which we define the proto-Milky Way to form, to compare the timing of disk settling with the broader assembly history of each system.

We quantify the linear association using the Pearson correlation coefficient,
\[
r = 
\frac{\sum_i (x_i-\bar{x})(y_i-\bar{y})}
{\sqrt{\sum_i (x_i-\bar{x})^2}\sqrt{\sum_i (y_i-\bar{y})^2}},
\]
where $x_i$ is the disk-settling lookback time, $y_i$ is the diskiness fraction, and bars denote sample means. The coefficient ranges from $-1$ to $1$: positive values indicate that the two quantities increase together, negative values indicate that one quantity decreases as the other increases, and values near zero indicate little linear association. In this case, the positive value of $r$ indicates that galaxies with larger disk-settling lookback times tend to have larger diskiness fractions. However, because our sample contains only a small number of simulations, we use $r$ as a descriptive measure of the trend rather than as a formal statistical test.

We find a moderate positive trend between diskiness fraction and $t_{\rm lb,form}$, suggesting that systems with larger disk-settling lookback times may be more likely to retain or develop coherent prograde rotation among old inner-galaxy stars. However, because our circularities are measured at $z=0$, this trend does not necessarily imply that these stars formed in a settled disk. Disk formation is gradual rather than a sharp transition, and coherent rotation may begin to emerge before the galaxy forms a dynamically cold disk. This is consistent with \citet{Elbadry2018}, who note that disks in their FIRE galaxies form around $z\sim1$, while also emphasizing that the onset of disk formation is gradual and varies across simulations. Thus, the prograde excess in old stars could reflect stars that formed with a preferred angular momentum direction in an early, still dynamically hot proto-disk, or stars that were later torqued into alignment with the disk.

This interpretation is consistent with previous FIRE-2 studies showing that disk formation is gradual. \citet{Yu2023} describe a bursty-disk ``spin-up'' phase in which newly formed stars begin moving onto more circular orbits before the fully settled thin disk forms. Similarly, Figure~1 of \citet{McCluskey2024} shows that stars formed during the Pre-Disk Era can have larger present-day azimuthal velocities than they had at formation, indicating that some old stars are later torqued onto more coherent rotational orbits as the disk forms and settles.

\section{Results and Discussion}
\label{res}
In this section, we explore the orbital and spatial properties of old inner galaxy stars in Milky Way–mass galaxies using the FIRE-2 simulations. We focus on three key questions: (1) Are old inner galaxy stars more likely to have been accreted or formed within the host galaxy? (2) Are accreted stars dynamically distinct from stars formed within the host galaxy? (3) What fraction of old inner galaxy stars' orbits are disk-like?

\subsection{Old Inner Galaxy Stars Predominantly Form in the Host Galaxy, but Accretion Ratio Increases with Age} 
\label{sec:exsitu}

\begin{figure*}
    \centering
    \includegraphics[width=\linewidth]{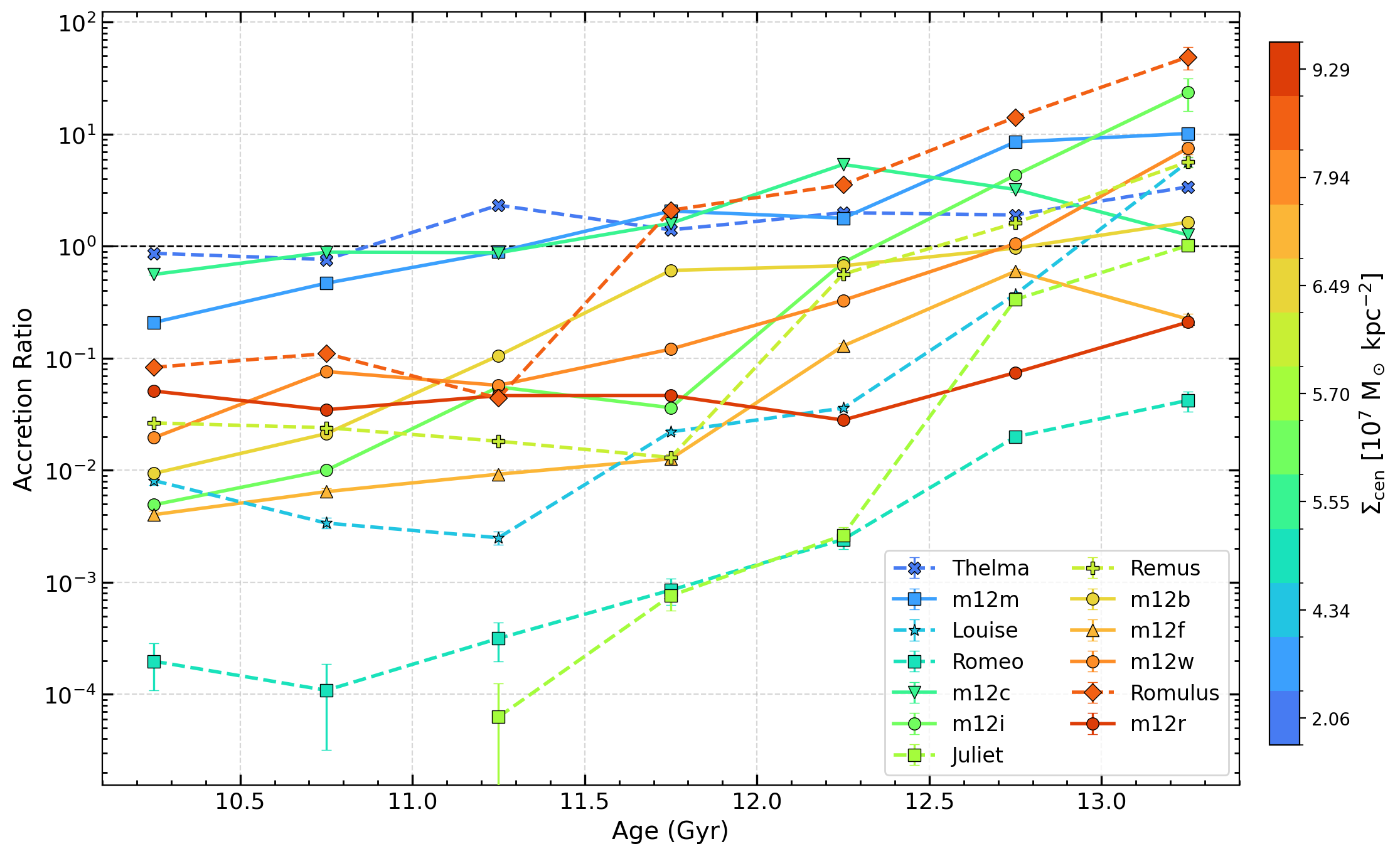}
    \caption{Accreted stellar ratio as a function of age for each FIRE-2 galaxy, with Poisson error bars. We define the accretion ratio as the number ratio of accreted to non-accreted stars within the old inner galaxy selection. Accreted stars are identified using particle-tracking catalogs (Section~\ref{sec:fire}). The black dashed line marks a ratio of unity, indicating equal contributions from stars formed within the host galaxy and accreted stars. Poisson uncertainties are calculated as $\sigma_R = R \sqrt{1/N_{\mathrm{ex}} + 1/N_{\mathrm{in}}}$, where $R$ is the fraction of accreted to non-accreted stars, and $N_{\mathrm{ex}}$ and $N_{\mathrm{in}}$ are the number of accreted and non-accreted stars in each age bin, respectively. Line colors encode the central stellar surface density $\Sigma_{\mathrm{cen}}$ (in units of $10^7\,M_\odot\,\mathrm{kpc}^{-2}$), allowing visual comparison between central concentration and accretion history. }

    \label{fig:dformratio}
\end{figure*}

To quantify the relative contributions of accretion versus star formation within the main galaxy across age, Figure~\ref{fig:dformratio} shows the ratio of accreted stars to those formed in the host galaxy as a function of stellar age. The accretion ratio was computed as the ratio of accreted stars to stars formed in the host galaxy, using accreted star indices identified from the particle-tracking catalogs described in Section~\ref{sec:fire} and applied to the stars in our old inner galaxy selection. Formation inside the main galaxy remains dominant across epochs in a majority of the FIRE-2 galaxies. Further, the downward trend in the accretion ratio with decreasing age suggests that formation within the host galaxy becomes increasingly dominant as age decreases.

Although most simulations do not show accretion-dominated growth across all stellar ages, many exhibit the highest ratio of accreted to in-situ stars at the oldest ages ($\gtrsim 12$ Gyr). In the oldest age bin, 10 of the 13 simulations exhibit accretion ratios greater than unity. This trend suggests that early epochs were the most active periods for accretion, even if in-galaxy star formation often remained the larger contributor during those times or overall. In the accreted population, we visually find that galaxies with higher central surface densities, $\Sigma_{\mathrm{cen}}$, tend to lie at higher diskiness values in the right panel of Figure~6, although we do not claim a strong correlation. This suggests that accreted material in these systems may be more likely to occupy aligned, disk-like orbits.

The total accretion ratio of old inner galaxy stars shows significant variation between the simulations (Table~\ref{tab:combined_table}), from less than 1\% in Romeo to greater than 50\% in Thelma and m12c. This halo diversity underscores that accretion is not uniformly dominant across Milky Way–mass galaxies in FIRE-2, but in some cases it accounts for a substantial portion of the old stellar population. These results highlight the interplay between sustained in-galaxy star formation and early accretion-driven growth as fundamental processes in the formation and evolution of Milky Way–like galaxies in FIRE-2.

Our finding that old inner galaxy stars predominantly formed in situ appears to be in tension with the results of \citet{Elbadry2018}, who find using FIRE simulations that the majority of the oldest stars ($z_{\rm form} > 5$, corresponding to a lookback time of $\sim 12.4$ Gyr) in the inner 10 kpc formed ex-situ and were subsequently accreted. However, we argue that this apparent tension is primarily driven by differences in the stellar age selection. \citet{Elbadry2018} focus exclusively on stars formed before $z = 5$, which selects a much earlier and more accretion-dominated epoch of galaxy assembly than our age $> 10$ Gyr threshold. Consistently with this interpretation, our Figure~\ref{fig:dformratio} shows that the accretion ratio in our sample exceeds unity in the majority of simulations at the oldest age bins ($\gtrsim 12$ Gyr), in agreement with the picture presented by \citet{Elbadry2018}. It is only at somewhat later times, corresponding to ages between 10 and 12 Gyr, that in-situ formation begins to dominate, pulling the overall accretion ratio below unity for most galaxies. Our results are therefore consistent with a picture in which the very earliest epoch of galaxy assembly ($z \gtrsim 5$) is accretion-dominated, but in-situ star formation becomes increasingly dominant at slightly later times, such that for the majority of galaxies in our sample, in-situ stars constitute the dominant contribution to the old inner galaxy population (age $> 10$ Gyr) as a whole, consistent with the low overall accretion ratios reported in Table~\ref{tab:combined_table}. We note, however, that two galaxies in our sample (Thelma and m12c) have total accretion ratios exceeding 50 per cent, highlighting the diversity of assembly histories across Milky Way-mass galaxies and underscoring that the dominance of in-situ formation is not universal.

\subsection{Comparison of Orbital Properties Accreted and In-Situ Populations}

\begin{figure*}
    \centering
    \includegraphics[width=.85\linewidth]{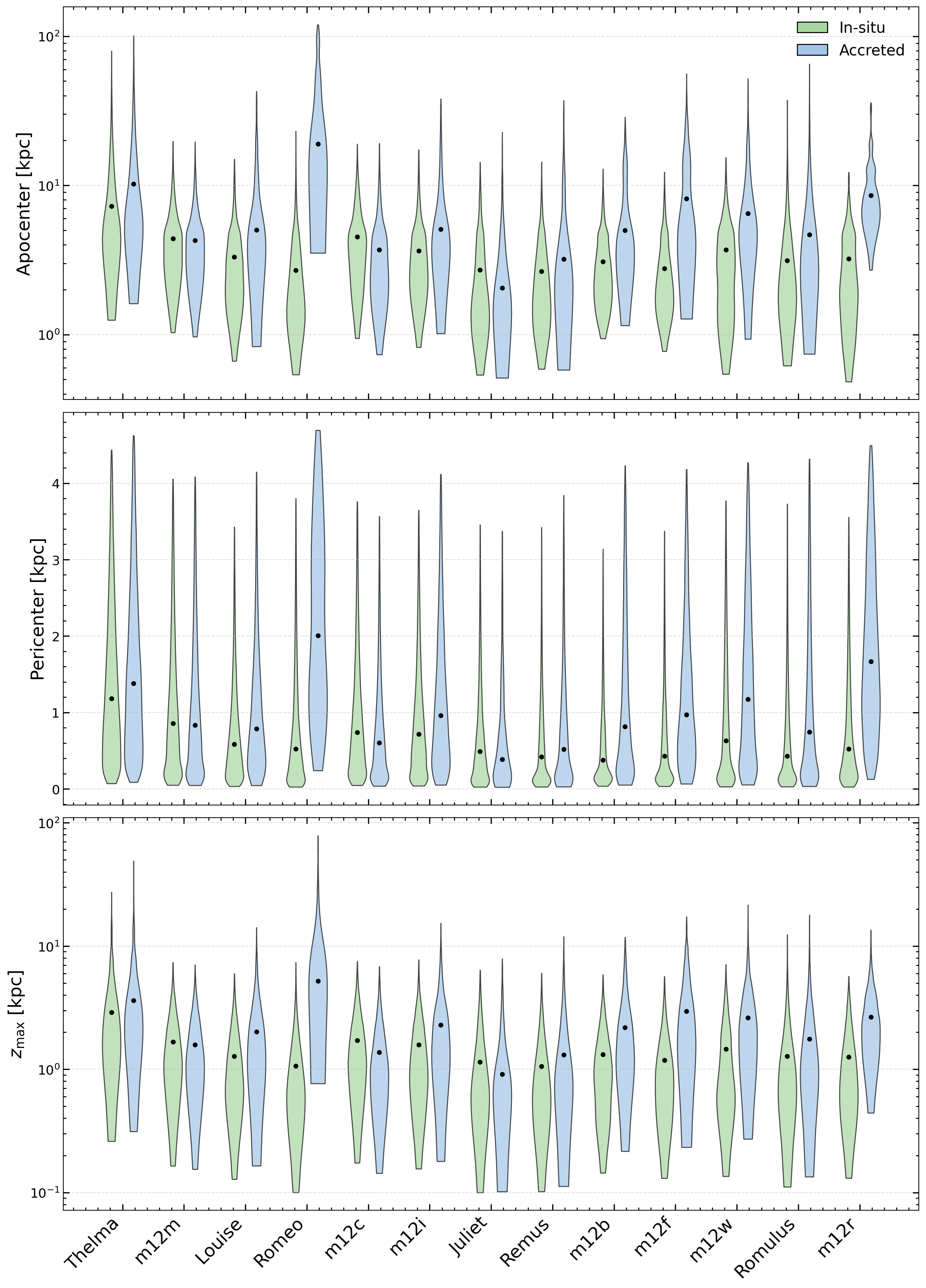}
    \caption{Violin distributions of apocenter, pericenter, and $z_{\mathrm{max}}$ for old inner galaxy stars in each FIRE-2 simulation, separated by formation origin using particle-tracking classifications. Green violins show stars formed within the host, while blue violins represent stars identified as accreted in the particle-tracking catalogs (Section~\ref{sec:fire}). Each violin illustrates the full distribution of orbital properties for that population, with the width indicating the relative density of stars and a black point marking the median. Across nearly all simulations, accreted stars exhibit larger apocenters, pericenters, and vertical excursions than in-situ stars, reflecting their typically more energetic and kinematically hot orbits. Simulations are sorted by increasing central stellar surface density ($\Sigma_{\mathrm{cen}}$).}

    \label{fig:box}
\end{figure*}

To further explore the orbital characteristics of old inner galaxy stars, we examine the distributions of their apocenter, pericenter, and $z_{\mathrm{max}}$ values in Figure~\ref{fig:box}. Each violin plot is mass-weighted and shows the full distribution shape of old inner galaxy stars separated into in-situ and accreted populations using the particle-tracking classifications. Across nearly all simulations, accreted stars (blue) exhibit significantly broader distributions and higher median values for apocenter, pericenter, and vertical height compared to their non-accreted counterparts (green). This distinction is especially visible in the apocenter and $z_{\mathrm{max}}$ panels, where accreted stars reach greater distances from the galactic center and plane of the disk. These trends likely reflect the more radial, eccentric, and vertically extended orbits of accreted stars, further supporting the conclusion that while in-situ star formation generally dominates galaxy growth, some old stellar populations in FIRE-2 galaxies are accreted and dynamically consistent with dispersion dominated kinematics. Although accreted stars generally exhibit hotter kinematics, many overlap with the kinematic properties of stars formed in the host galaxy. As a result, these statistics cannot reliably classify individual stars as accreted or not. Instead, they are more meaningful when applied to populations as a whole, highlighting differences between the accreted and in-situ groups.

\subsection{Old Inner Galaxy Stars are Largely Dispersion Dominated} 
\label{sec:halo}

We quantify the degree of ordered rotation among old inner galaxy stars using the diskiness fraction, which measures the excess of prograde over retrograde stellar mass. This metric, derived from orbital circularity histograms and defined in Section~\ref{sec:orbit}, provides a simple way to compare rotational support across simulations (Table~\ref{tab:combined_table}).

Most simulations produce low diskiness fractions (e.g., $<0.1$ for Thelma, m12i, Louise, and Juliet), consistent with dispersion supported kinematics. Some simulations, including m12m and Romeo, exhibit moderate disk fractions ($\sim$ 0.35), indicating that a subset of old inner galaxy stars may still retain some net rotation. Overall, this analysis supports the conclusion that old stellar populations in FIRE-2 galaxies are dispersion dominated, with little net rotational support.

We find no significant correlation between accretion ratio and diskiness fraction (Table~\ref{tab:combined_table}). The prevalence of accreted stars does not predict the degree of disk-like structure in the old stellar population of the inner galaxy. This indicates that accretion is only one of many factors shaping orbital structure, and that diverse evolutionary pathways can produce either disky or non-disky configurations.

\begin{figure*}
    \centering
    \includegraphics[width=\linewidth]{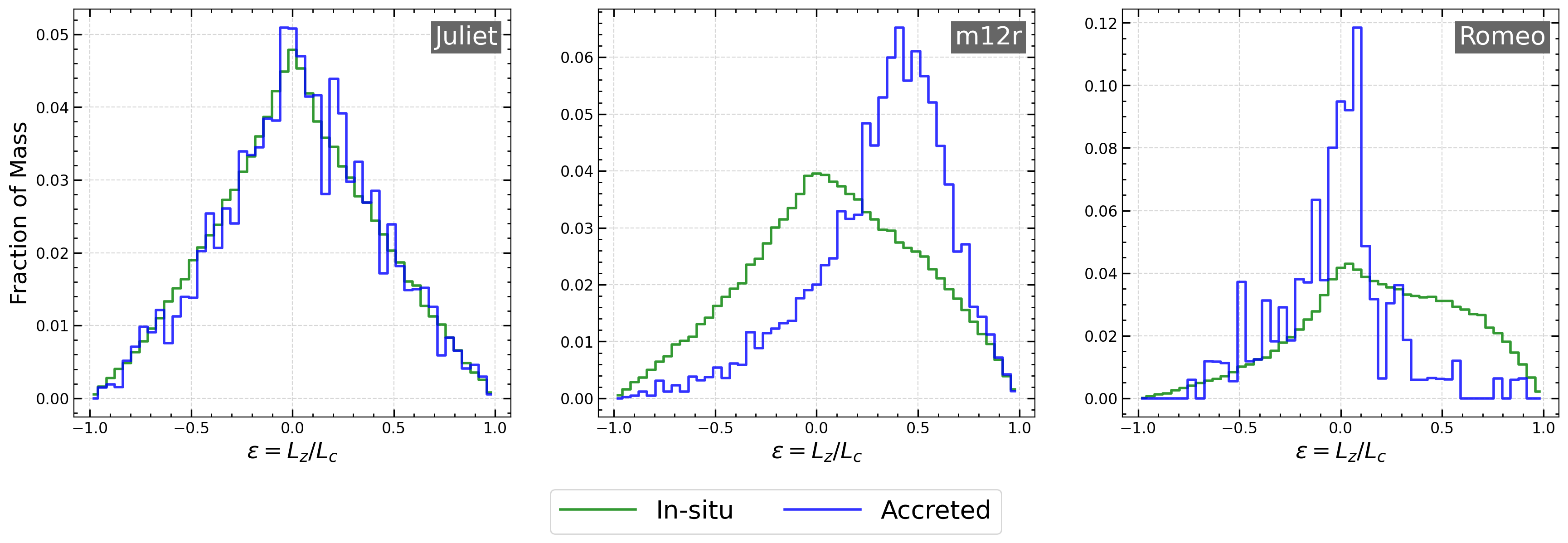}
    \caption{Mass-weighted distributions of orbital circularity ($\epsilon = L_z/L_c$) for old inner galaxy stars in three representative simulations (Juliet, m12r, and Romeo), separated using particle-tracking classifications. Green curves show stars born in the host, while blue curves show stars associated with accreted structures as indicated in the particle-tracking catalogs (Section~\ref{sec:fire}). Juliet exhibits a relatively symmetric distribution with similar shapes for both components, consistent with the absence of a prominent disk. In m12r, the accreted population shows a strong high-circularity feature produced by a coherent accretion event. Romeo, by contrast, displays a more disk-like in-situ skew, reflecting significant rotational support among its old stars. Together, these examples illustrate the diversity of dynamical histories present in the FIRE-2 suite and highlight how accretion and in-situ formation shape the circularity structure of the inner galaxy.}

    \label{fig:massdisk}
\end{figure*}

Figure~\ref{fig:massdisk} shows the normalized distribution of orbital circularity ($\epsilon = L_z / L_c$) for old inner galaxy stars, separated into in-situ and accreted populations using the particle-tracking classifications. This plot provides another view of the distinctions between the dynamics of these two populations. In some simulations (e.g., Juliet), the distributions of the in-situ and accreted populations appear broadly similar and are both peaked around $\epsilon \sim 0$, indicating a largely dispersion-supported population. However, in some cases, such as m12r, the accreted distribution skews toward the positive side. We see the same skew for in-situ stars in Romeo. Such differences illustrate how the balance of rotational order between stars formed in the host and accreted stars differs from galaxy to galaxy. Taken together, the distributions suggest that while most old inner galaxy stars are kinematically hot regardless of origin, the precise kinematic imprint of accretion and non-accretion based formation histories vary across galaxies. The complete set of simulation panels is provided in Figure~\ref{fig:eps_histos_all} of the appendix.

A prograde skew in the circularity distribution of in-situ stars is consistent with secular evolutionary processes that build up the disk over time. In contrast, a similar skew observed in accreted populations is less expected and may suggest that some accretion events occurred along preferred paths, such as dark matter filaments, causing the infalling stars to align kinematically with the host galaxy’s disk \citep{Arora2024a}.

\begin{figure*}
    \centering
    \includegraphics[width=\linewidth]{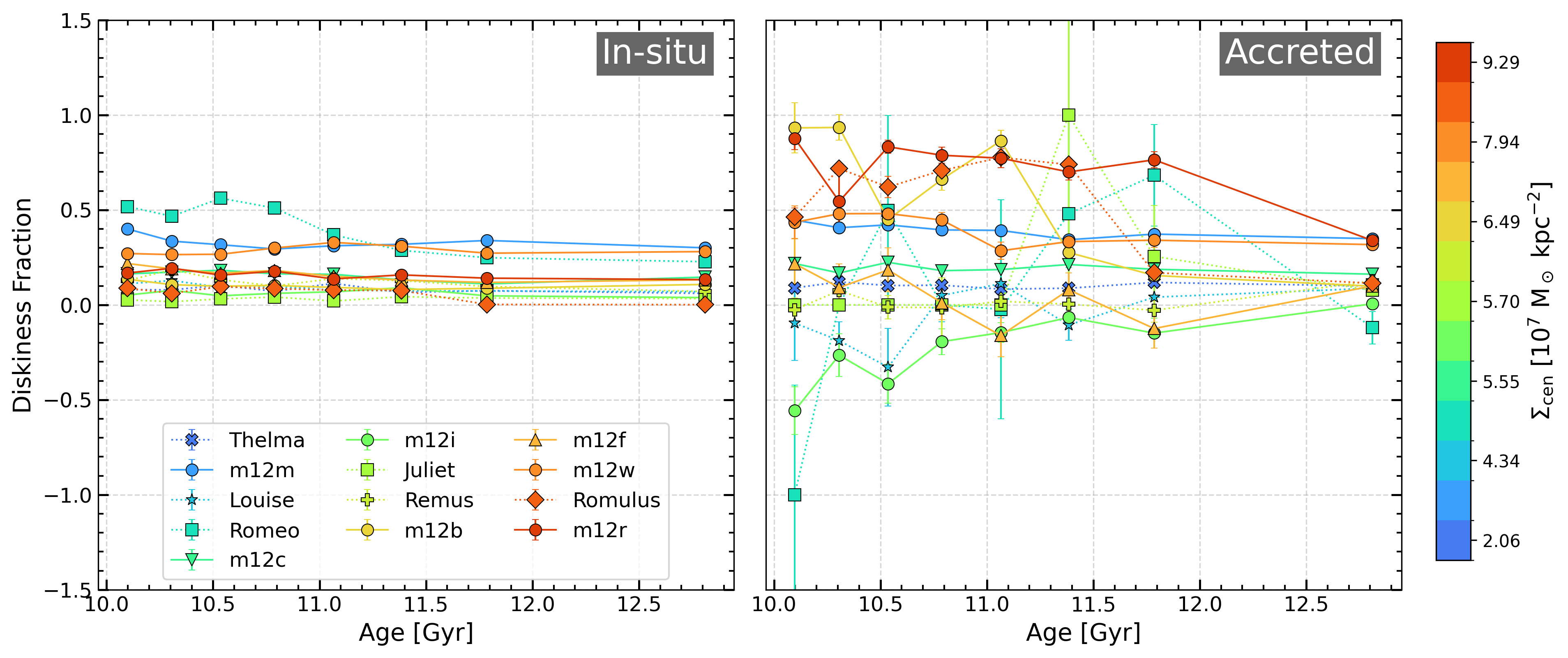}
    \caption{
    Diskiness fraction as a function of stellar age, shown separately for in-situ stars (left) and accreted stars (right) identified in the particle-tracking catalogs (Section~\ref{sec:fire}). The fraction of stars with disk-like dynamics is computed using the $L_z/L_c$ mass histograms (Figure~\ref{fig:flip}), where the difference in mass between the prograde side ($\epsilon > 0$) and the retrograde side ($\epsilon < 0$), normalized by the total stellar mass, reflects the degree of prograde rotation. Poisson error bars represent statistical uncertainties in each age bin and are approximated as $\sigma = 1/\sqrt{N}$, where $N$ is the number of stars in that bin. Simulations exhibit a wide range of behaviors: some (e.g., m12i, Louise) show low or even negative disk fractions for their accreted populations, while others retain moderate rotational signatures across age. The color gradient reflects the central stellar surface density ($\Sigma_{\mathrm{cen}}$) of each simulation.
}

    \label{fig:3disk}
\end{figure*}

In Figure~\ref{fig:3disk}, we explore how diskiness  varies with stellar age and origin. We plot the diskiness fraction as a function of age for in-situ stars (left panel), and accreted stars (right panel). These classifications are taken from the particle-tracking catalogs (Section~\ref{sec:fire}). Poisson error bars are included. Across simulations, in-situ stars show relatively flat diskiness with age, indicating that the rotational support of the in-situ component remains fairly stable. However, some simulations (e.g. Romeo and m12m) exhibit higher diskiness at younger ages within the in-situ population. In contrast, the accreted population spans a much broader range of behaviors. We find that galaxies with higher $\Sigma_{\mathrm{cen}}$ tend to exhibit greater diskiness among their accreted stars. These differences underscore how the rotational structure of old inner galaxy stars depends not only on formation history, but also on the timing and nature of accretion.

\section{Conclusion}
\label{sec:conc}

In this work, we analyzed the orbital properties of old (age $>10$ Gyr) inner galaxy ($r_\mathrm{GC}$ $<5$ kpc) stars in thirteen FIRE-2 Milky Way-mass galaxies and compared accreted and non-accreted stellar populations. This analysis allows us to assess the relative contributions of hierarchical accretion and secular processes in shaping the early dynamical structure of Milky Way-mass galaxies. Using orbital circularity as a diagnostic, we quantified the prominence of disk-like dynamics across simulations and found that most old inner galaxy stars are dynamically hot and dispersion-supported. While this trend holds broadly, a subset of galaxies, such as m12w, m12m, and Romeo exhibit disk-like dynamics, indicating residual rotational support in their accreted or non-accreted components.

Using particle-tracking formation classifications to separate in-situ and accreted stars, our comparison of their circularity distributions demonstrates how varying assembly histories leave distinct imprints on stellar kinematics. Some galaxies show near-identical accreted and non-accreted distributions, while others show significant asymmetry, emphasizing the diversity of inner galaxy formation pathways. Additionally, our analysis of diskiness as a function of age shows that the in-situ population remains relatively stable across age, while the accreted population exhibits substantially more scatter and variation in rotational support. This may be due to a correlation between the disk and accretion directions, and the possibility of accreted material retaining angular momentum from infall along dark matter filaments, leading to alignment with the galactic disk \citep{Arora2024a}.

Our analysis supports several key conclusions regarding the origin and dynamical properties of old inner galaxy stars in Milky Way-mass galaxies:
\begin{enumerate}
    \item In-situ star formation dominates the old inner galaxy population in most galaxies overall, but at the earliest times the accretion ratio is highest, with a majority of systems undergoing a temporary phase of accretion-dominated growth.
    \item Old inner galaxy stars show kinematically hot orbits, with accreted populations extending to larger apocenters, pericenters, and $z_{\mathrm{max}}$ values than the in-situ population. However, the in-situ and accreted populations exhibit substantial overlap, indicating that kinematics alone cannot reliably identify the accretion history of individual stars.
    \item The diskiness fraction spans a moderate range within the in-situ population, from $\sim$ 0\% to 50\%, and remains largely constant with stellar age in each simulation. In contrast, the accreted population exhibits a much wider spread and substantially greater variation in diskiness across age. The total diskiness fraction ranges from 3.4\% to 37.6\% across simulations.
\end{enumerate}

Looking ahead, this work provides theoretical predictions for the kinematic structure and origin of old inner galaxy stars that can be tested against current and upcoming Milky Way observations. Surveys like SDSS-V, which will map stellar kinematics and chemical abundances across the Milky Way with unprecedented coverage and precision \citep{Chandra2025}, offer an opportunity to directly compare our simulation-based results to real data. In particular, our findings regarding the orbital properties and disk-like dynamics of accreted stars may help interpret observations of old populations in the bulge and inner galaxy. As more high-resolution spectroscopic and positional data become available, it will be possible to further constrain the formation pathways of the Milky Way’s oldest stars and assess the role of accretion and internal formation processes in shaping the Galaxy's present-day structure.

\section*{Acknowledgements}

 This material is based upon work supported by the National Science Foundation under Award No. 2303831. RES was also supported in this project by Sloan Foundation award FG-2023-20669 and Simons Foundation grant 1018462. LN is supported by the Sloan Fellowship, the NSF CAREER award 2337864, and the NSF award 2307788. 

\section*{Data Availability}
All of the simulations underlying this work are publicly available at \url{https://flathub.flatironinstitute.org/fire}.

\bibliography{bibliography}{}

@ARTICLE{Arora2024a,
       author = {{Arora}, Arpit and {Garavito-Camargo}, Nicol{\'a}s and {Sanderson}, Robyn E. and {Cunningham}, Emily C. and {Wetzel}, Andrew and {Panithanpaisal}, Nondh and {Barry}, Megan},
        title = "{LMC-driven Anisotropic Boosts in Stream{\textendash}Subhalo Interactions}",
      journal = {\apj},
         year = 2024,
        month = oct,
       volume = {974},
       number = {2},
          eid = {286},
        pages = {286},
          doi = {10.3847/1538-4357/ad7375},
archivePrefix = {arXiv},
       eprint = {2309.15998},
 primaryClass = {astro-ph.GA},
       adsurl = {https://ui.adsabs.harvard.edu/abs/2024ApJ...974..286A}
}

@ARTICLE{Arora2022,
       author = {{Arora}, Arpit and {Sanderson}, Robyn E. and {Panithanpaisal}, Nondh and {Cunningham}, Emily C. and {Wetzel}, Andrew and {Garavito-Camargo}, Nicol{\'a}s},
        title = "{On the Stability of Tidal Streams in Action Space}",
      journal = {\apj},
         year = 2022,
        month = nov,
       volume = {939},
       number = {1},
          eid = {2},
        pages = {2},
          doi = {10.3847/1538-4357/ac93fb},
archivePrefix = {arXiv},
       eprint = {2207.13481},
 primaryClass = {astro-ph.GA},
       adsurl = {https://ui.adsabs.harvard.edu/abs/2022ApJ...939....2A}
}

@article{Arora2024b,
   title={Efficient and Accurate Force Replay in Cosmological-baryonic Simulations},
   volume={977},
   ISSN={1538-4357},
   url={http://dx.doi.org/10.3847/1538-4357/ad88f0},
   DOI={10.3847/1538-4357/ad88f0},
   number={1},
   journal={The Astrophysical Journal},
   publisher={American Astronomical Society},
   author={Arora, Arpit and Sanderson, Robyn and Regan, Christopher and Garavito-Camargo, Nicolás and Bregou, Emily and Panithanpaisal, Nondh and Wetzel, Andrew and Cunningham, Emily C. and Loebman, Sarah R. and Dropulic, Adriana and Shipp, Nora},
   year={2024},
   month=Nov, pages={23} }

@ARTICLE{Lucey2021,
       author = {{Lucey}, Madeline and {Hawkins}, Keith and {Ness}, Melissa and {Debattista}, Victor P. and {Luna}, Alice and {Asplund}, Martin and {Bensby}, Thomas and {Casagrande}, Luca and {Feltzing}, Sofia and {Freeman}, Kenneth C. and {Kobayashi}, Chiaki and {Marino}, Anna F.},
        title = "{The COMBS Survey - II. Distinguishing the metal-poor bulge from the halo interlopers}",
      journal = {\mnras},
         year = 2021,
        month = mar,
       volume = {501},
       number = {4},
        pages = {5981-5996},
          doi = {10.1093/mnras/stab003},
archivePrefix = {arXiv},
       eprint = {2009.03886},
 primaryClass = {astro-ph.GA},
       adsurl = {https://ui.adsabs.harvard.edu/abs/2021MNRAS.501.5981L}
}

@ARTICLE{Elbadry2018,
   author = {{El-Badry}, K. and {Bland-Hawthorn}, J. and {Wetzel}, A. and 
	{Quataert}, E. and {Weisz}, D.~R. and {Boylan-Kolchin}, M. and 
	{Hopkins}, P.~F. and {Faucher-Gigu{\`e}re}, C.-A. and {Kere{\v s}}, D. and 
	{Garrison-Kimmel}, S.},
    title = "{Where are the most ancient stars in the Milky Way?}",
  journal = {\mnras},
archivePrefix = "arXiv",
   eprint = {1804.00659},
     year = 2018,
    month = oct,
   volume = 480,
    pages = {652-668},
      doi = {10.1093/mnras/sty1864},
   adsurl = {http://adsabs.harvard.edu/abs/2018MNRAS.480..652E}
}

@dataset{Vasiliev2021,
       author = {{Vasiliev}, E. and {Baumgardt}, H.},
        title = "{VizieR Online Data Catalog: Gaia EDR3 view on Galactic globular clusters (Vasiliev+, 2021)}",
 howpublished = {VizieR On-line Data Catalog: J/MNRAS/505/5978. Originally published in: 2021MNRAS.505.5978V},
         year = 2021,
        month = jul,
          eid = {J/MNRAS/505/5978},
       adsurl = {https://ui.adsabs.harvard.edu/abs/2021yCat..75055978V}
}

@article{White1978,
	Adsurl = {http://adsabs.harvard.edu/abs/1978MNRAS.183..341W},
	Author = {{White}, S.~D.~M. and {Rees}, M.~J.},
	Doi = {10.1093/mnras/183.3.341},
	Journal = {\mnras},
	Month = may,
	Pages = {341-358},
	Title = {{Core condensation in heavy halos - A two-stage theory for galaxy formation and clustering}},
	Volume = 183,
	Year = 1978}

@article{Freeman2002,
	Adsurl = {http://adsabs.harvard.edu/abs/2002ARA%26A..40..487F},
	Author = {{Freeman}, K. and {Bland-Hawthorn}, J.},
	Doi = {10.1146/annurev.astro.40.060401.093840},
	Eprint = {astro-ph/0208106},
	Journal = {\araa},
	Pages = {487-537},
	Title = {{The New Galaxy: Signatures of Its Formation}},
	Volume = 40,
	Year = 2002}

@article{Cooper2010,
	Adsurl = {http://adsabs.harvard.edu/abs/2010MNRAS.406..744C},
	Archiveprefix = {arXiv},
	Author = {{Cooper}, A.~P. and {Cole}, S. and {Frenk}, C.~S. and {White}, S.~D.~M. and {Helly}, J. and {Benson}, A.~J. and {De Lucia}, G. and {Helmi}, A. and {Jenkins}, A. and {Navarro}, J.~F. and {Springel}, V. and {Wang}, J.},
	Doi = {10.1111/j.1365-2966.2010.16740.x},
	Eprint = {0910.3211},
	Journal = {\mnras},
	Month = aug,
	Pages = {744-766},
	Primaryclass = {astro-ph.GA},
	Title = {{Galactic stellar haloes in the CDM model}},
	Volume = 406,
	Year = 2010}

@article{Bullock2005,
	Adsurl = {http://adsabs.harvard.edu/abs/2005ApJ...635..931B},
	Author = {{Bullock}, J.~S. and {Johnston}, K.~V.},
	Doi = {10.1086/497422},
	Eprint = {astro-ph/0506467},
	Journal = {\apj},
	Month = dec,
	Pages = {931-949},
	Title = {{Tracing Galaxy Formation with Stellar Halos. I. Methods}},
	Volume = 635,
	Year = 2005}

@article{Tissera2013,
	Adsurl = {http://adsabs.harvard.edu/abs/2013MNRAS.432.3391T},
	Archiveprefix = {arXiv},
	Author = {{Tissera}, P.~B. and {Scannapieco}, C. and {Beers}, T.~C. and {Carollo}, D.},
	Doi = {10.1093/mnras/stt691},
	Eprint = {1301.1301},
	Journal = {\mnras},
	Month = jul,
	Pages = {3391-3400},
	Primaryclass = {astro-ph.GA},
	Title = {{Stellar haloes of simulated Milky-Way-like galaxies: chemical and kinematic properties}},
	Volume = 432,
	Year = 2013}

@article{Purcell2010,
	Adsurl = {http://adsabs.harvard.edu/abs/2010MNRAS.404.1711P},
	Archiveprefix = {arXiv},
	Author = {{Purcell}, C.~W. and {Bullock}, J.~S. and {Kazantzidis}, S.},
	Doi = {10.1111/j.1365-2966.2010.16429.x},
	Eprint = {0910.5481},
	Journal = {\mnras},
	Month = jun,
	Pages = {1711-1718},
	Primaryclass = {astro-ph.GA},
	Title = {{Heated disc stars in the stellar halo}},
	Volume = 404,
	Year = 2010}

@article{Font2011,
	Adsurl = {http://adsabs.harvard.edu/abs/2011MNRAS.416.2802F},
	Archiveprefix = {arXiv},
	Author = {{Font}, A.~S. and {McCarthy}, I.~G. and {Crain}, R.~A. and {Theuns}, T. and {Schaye}, J. and {Wiersma}, R.~P.~C. and {Dalla Vecchia}, C.},
	Doi = {10.1111/j.1365-2966.2011.19227.x},
	Eprint = {1102.2526},
	Journal = {\mnras},
	Month = oct,
	Pages = {2802-2820},
	Primaryclass = {astro-ph.CO},
	Title = {{Cosmological simulations of the formation of the stellar haloes around disc galaxies}},
	Volume = 416,
	Year = 2011}

@article{Font2006a,
	Adsurl = {http://adsabs.harvard.edu/abs/2006ApJ...638..585F},
	Author = {{Font}, A.~S. and {Johnston}, K.~V. and {Bullock}, J.~S. and {Robertson}, B.~E.},
	Doi = {10.1086/498970},
	Eprint = {astro-ph/0507114},
	Journal = {\apj},
	Month = feb,
	Pages = {585-595},
	Title = {{Chemical Abundance Distributions of Galactic Halos and Their Satellite Systems in a {$\Lambda$}CDM Universe}},
	Volume = 638,
	Year = 2006}

@article{Robertson2005,
	Adsurl = {http://adsabs.harvard.edu/abs/2005ApJ...632..872R},
	Author = {{Robertson}, B. and {Bullock}, J.~S. and {Font}, A.~S. and {Johnston}, K.~V. and {Hernquist}, L.},
	Doi = {10.1086/452619},
	Eprint = {astro-ph/0501398},
	Journal = {\apj},
	Month = oct,
	Pages = {872-881},
	Title = {{{$\Lambda$} Cold Dark Matter, Stellar Feedback, and the Galactic Halo Abundance Pattern}},
	Volume = 632,
	Year = 2005}

@article{Helmi2008,
	Adsurl = {http://adsabs.harvard.edu/abs/2008A%26ARv..15..145H},
	Archiveprefix = {arXiv},
	Author = {{Helmi}, A.},
	Doi = {10.1007/s00159-008-0009-6},
	Eprint = {0804.0019},
	Journal = {\aapr},
	Month = jun,
	Pages = {145-188},
	Title = {{The stellar halo of the Galaxy}},
	Volume = 15,
	Year = 2008}

@article{Zolotov2009,
	Adsurl = {http://adsabs.harvard.edu/abs/2009ApJ...702.1058Z},
	Archiveprefix = {arXiv},
	Author = {{Zolotov}, A. and {Willman}, B. and {Brooks}, A.~M. and {Governato}, F. and {Brook}, C.~B. and {Hogg}, D.~W. and {Quinn}, T. and {Stinson}, G.},
	Doi = {10.1088/0004-637X/702/2/1058},
	Eprint = {0904.3333},
	Journal = {\apj},
	Month = sep,
	Pages = {1058-1067},
	Primaryclass = {astro-ph.GA},
	Title = {{The Dual Origin of Stellar Halos}},
	Volume = 702,
	Year = 2009}

@article{Brown2014,
	Adsurl = {http://adsabs.harvard.edu/abs/2014arXiv1401.7342B},
	Archiveprefix = {arXiv},
	Author = {{Brown}, W.~R. and {Geller}, M.~J. and {Kenyon}, S.~J.},
	Eprint = {1401.7342},
	Journal = {ArXiv e-prints},
	Month = jan,
	Primaryclass = {astro-ph.SR},
	Title = {{MMT Hypervelocity Star Survey III: A Complete Survey of Faint B-type Stars in the Northern Milky Way Halo}},
	Volume = {arXiv:1401.7342},
	Year = 2014}

@article{Carollo2007,
	Adsurl = {http://adsabs.harvard.edu/abs/2007Natur.450.1020C},
	Archiveprefix = {arXiv},
	Author = {{Carollo}, D. and {Beers}, T.~C. and {Lee}, Y.~S. and {Chiba}, M. and {Norris}, J.~E. and {Wilhelm}, R. and {Sivarani}, T. and {Marsteller}, B. and {Munn}, J.~A. and {Bailer-Jones}, C.~A.~L. and {Fiorentin}, P.~R. and {York}, D.~G.},
	Doi = {10.1038/nature06460},
	Eprint = {0706.3005},
	Journal = {\nat},
	Month = dec,
	Pages = {1020-1025},
	Title = {{Two stellar components in the halo of the Milky Way}},
	Volume = 450,
	Year = 2007}

@ARTICLE{Tumlinson2010,
       author = {{Tumlinson}, Jason},
        title = "{Chemical Evolution in Hierarchical Models of Cosmic Structure. II. The Formation of the Milky Way Stellar Halo and the Distribution of the Oldest Stars}",
      journal = {\apj},
         year = 2010,
        month = Jan,
       volume = {708},
        pages = {1398-1418},
          doi = {10.1088/0004-637X/708/2/1398},
archivePrefix = {arXiv},
       eprint = {0911.1786},
 primaryClass = {astro-ph.GA},
       adsurl = {https://ui.adsabs.harvard.edu/\#abs/2010ApJ...708.1398T}
}

@ARTICLE{Helmi2018,
       author = {{Helmi}, Amina and {Babusiaux}, Carine and {Koppelman}, Helmer H. and
        {Massari}, Davide and {Veljanoski}, Jovan and {Brown}, Anthony
        G.~A.},
        title = "{The merger that led to the formation of the Milky Way's inner stellar halo and thick disk}",
      journal = {\nat},
         year = 2018,
        month = Nov,
       volume = {563},
        pages = {85-88},
          doi = {10.1038/s41586-018-0625-x},
archivePrefix = {arXiv},
       eprint = {1806.06038},
 primaryClass = {astro-ph.GA},
       adsurl = {https://ui.adsabs.harvard.edu/\#abs/2018Natur.563...85H}
}

@ARTICLE{El-Badry2018a,
       author = {{El-Badry}, Kareem and {Rix}, Hans-Walter and {Ting}, Yuan-Sen and {Weisz}, Daniel R. and {Bergemann}, Maria and {Cargile}, Phillip and {Conroy}, Charlie and {Eilers}, Anna-Christina},
        title = "{Signatures of unresolved binaries in stellar spectra: implications for spectral fitting}",
      journal = {\mnras},
         year = 2018,
        month = feb,
       volume = {473},
       number = {4},
        pages = {5043-5049},
          doi = {10.1093/mnras/stx2758},
archivePrefix = {arXiv},
       eprint = {1709.03983},
 primaryClass = {astro-ph.SR},
       adsurl = {https://ui.adsabs.harvard.edu/abs/2018MNRAS.473.5043E}
}

@ARTICLE{Santistevan2020,
       author = {{Santistevan}, Isaiah B. and {Wetzel}, Andrew and {El-Badry}, Kareem and {Bland-Hawthorn}, Joss and {Boylan-Kolchin}, Michael and {Bailin}, Jeremy and {Faucher-Gigu{\`e}re}, Claude-Andr{\'e} and {Benincasa}, Samantha},
        title = "{The formation times and building blocks of Milky Way-mass galaxies in the FIRE simulations}",
      journal = {\mnras},
         year = 2020,
        month = sep,
       volume = {497},
       number = {1},
        pages = {747-764},
          doi = {10.1093/mnras/staa1923},
archivePrefix = {arXiv},
       eprint = {2001.03178},
 primaryClass = {astro-ph.GA},
       adsurl = {https://ui.adsabs.harvard.edu/abs/2020MNRAS.497..747S}
}

@ARTICLE{Beers2005,
       author = {{Beers}, Timothy C. and {Christlieb}, Norbert},
        title = "{The Discovery and Analysis of Very Metal-Poor Stars in the Galaxy}",
      journal = {\araa},
         year = 2005,
        month = sep,
       volume = {43},
       number = {1},
        pages = {531-580},
          doi = {10.1146/annurev.astro.42.053102.134057},
       adsurl = {https://ui.adsabs.harvard.edu/abs/2005ARA&A..43..531B}
}

@ARTICLE{Frebel2015,
       author = {{Frebel}, Anna and {Norris}, John E.},
        title = "{Near-Field Cosmology with Extremely Metal-Poor Stars}",
      journal = {\araa},
         year = 2015,
        month = aug,
       volume = {53},
        pages = {631-688},
          doi = {10.1146/annurev-astro-082214-122423},
archivePrefix = {arXiv},
       eprint = {1501.06921},
 primaryClass = {astro-ph.SR},
       adsurl = {https://ui.adsabs.harvard.edu/abs/2015ARA&A..53..631F}
}

@ARTICLE{Wetzel2016,
       author = {{Wetzel}, Andrew R. and {Hopkins}, Philip F. and {Kim}, Ji-hoon and {Faucher-Gigu{\`e}re}, Claude-Andr{\'e} and {Kere{\v{s}}}, Du{\v{s}}an and {Quataert}, Eliot},
        title = "{Reconciling Dwarf Galaxies with {\ensuremath{\Lambda}}CDM Cosmology: Simulating a Realistic Population of Satellites around a Milky Way-mass Galaxy}",
      journal = {\apjl},
         year = 2016,
        month = aug,
       volume = {827},
       number = {2},
          eid = {L23},
        pages = {L23},
          doi = {10.3847/2041-8205/827/2/L23},
archivePrefix = {arXiv},
       eprint = {1602.05957},
 primaryClass = {astro-ph.GA},
       adsurl = {https://ui.adsabs.harvard.edu/abs/2016ApJ...827L..23W}
}

@ARTICLE{Garrison-Kimmel2019,
       author = {{Garrison-Kimmel}, Shea and {Hopkins}, Philip F. and {Wetzel}, Andrew and {Bullock}, James S. and {Boylan-Kolchin}, Michael and {Kere{\v{s}}}, Du{\v{s}}an and {Faucher-Gigu{\`e}re}, Claude-Andr{\'e} and {El-Badry}, Kareem and {Lamberts}, Astrid and {Quataert}, Eliot and {Sanderson}, Robyn},
        title = "{The Local Group on FIRE: dwarf galaxy populations across a suite of hydrodynamic simulations}",
      journal = {\mnras},
         year = 2019,
        month = jul,
       volume = {487},
       number = {1},
        pages = {1380-1399},
          doi = {10.1093/mnras/stz1317},
archivePrefix = {arXiv},
       eprint = {1806.04143},
 primaryClass = {astro-ph.GA},
       adsurl = {https://ui.adsabs.harvard.edu/abs/2019MNRAS.487.1380G}
}

@ARTICLE{Hopkins2018b,
       author = {{Hopkins}, Philip F. and {Wetzel}, Andrew and {Kere{\v{s}}}, Du{\v{s}}an and {Faucher-Gigu{\`e}re}, Claude-Andr{\'e} and {Quataert}, Eliot and {Boylan-Kolchin}, Michael and {Murray}, Norman and {Hayward}, Christopher C. and {Garrison-Kimmel}, Shea and {Hummels}, Cameron and {Feldmann}, Robert and {Torrey}, Paul and {Ma}, Xiangcheng and {Angl{\'e}s-Alc{\'a}zar}, Daniel and {Su}, Kung-Yi and {Orr}, Matthew and {Schmitz}, Denise and {Escala}, Ivanna and {Sanderson}, Robyn and {Grudi{\'c}}, Michael Y. and {Hafen}, Zachary and {Kim}, Ji-Hoon and {Fitts}, Alex and {Bullock}, James S. and {Wheeler}, Coral and {Chan}, T.~K. and {Elbert}, Oliver D. and {Narayanan}, Desika},
        title = "{FIRE-2 simulations: physics versus numerics in galaxy formation}",
      journal = {\mnras},
         year = 2018,
        month = oct,
       volume = {480},
       number = {1},
        pages = {800-863},
          doi = {10.1093/mnras/sty1690},
archivePrefix = {arXiv},
       eprint = {1702.06148},
 primaryClass = {astro-ph.GA},
       adsurl = {https://ui.adsabs.harvard.edu/abs/2018MNRAS.480..800H}
}

@ARTICLE{Hopkins2015,
       author = {{Hopkins}, Philip F.},
        title = "{A new class of accurate, mesh-free hydrodynamic simulation methods}",
      journal = {\mnras},
         year = 2015,
        month = jun,
       volume = {450},
       number = {1},
        pages = {53-110},
          doi = {10.1093/mnras/stv195},
archivePrefix = {arXiv},
       eprint = {1409.7395},
 primaryClass = {astro-ph.CO},
       adsurl = {https://ui.adsabs.harvard.edu/abs/2015MNRAS.450...53H}
}

@ARTICLE{Sanderson2018,
       author = {{Sanderson}, Robyn E. and {Garrison-Kimmel}, Shea and {Wetzel}, Andrew and {Keung Chan}, Tsang and {Hopkins}, Philip F. and {Kere{\v{s}}}, Du{\v{s}}an and {Escala}, Ivanna and {Faucher-Gigu{\`e}re}, Claude-Andr{\'e} and {Ma}, Xiangcheng},
        title = "{Reconciling Observed and Simulated Stellar Halo Masses}",
      journal = {\apj},
         year = 2018,
        month = dec,
       volume = {869},
       number = {1},
          eid = {12},
        pages = {12},
          doi = {10.3847/1538-4357/aaeb33},
archivePrefix = {arXiv},
       eprint = {1712.05808},
 primaryClass = {astro-ph.GA},
       adsurl = {https://ui.adsabs.harvard.edu/abs/2018ApJ...869...12S}
}

@ARTICLE{Planck2020,
       author = {{Planck Collaboration} and {Aghanim}, N. and {Akrami}, Y. and {Ashdown}, M. and {Aumont}, J. and {Baccigalupi}, C. and {Ballardini}, M. and {Banday}, A.~J. and {Barreiro}, R.~B. and {Bartolo}, N. and {Basak}, S. and {Battye}, R. and {Benabed}, K. and {Bernard}, J. -P. and {Bersanelli}, M. and {Bielewicz}, P. and {Bock}, J.~J. and {Bond}, J.~R. and {Borrill}, J. and {Bouchet}, F.~R. and {Boulanger}, F. and {Bucher}, M. and {Burigana}, C. and {Butler}, R.~C. and {Calabrese}, E. and {Cardoso}, J. -F. and {Carron}, J. and {Challinor}, A. and {Chiang}, H.~C. and {Chluba}, J. and {Colombo}, L.~P.~L. and {Combet}, C. and {Contreras}, D. and {Crill}, B.~P. and {Cuttaia}, F. and {de Bernardis}, P. and {de Zotti}, G. and {Delabrouille}, J. and {Delouis}, J. -M. and {Di Valentino}, E. and {Diego}, J.~M. and {Dor{\'e}}, O. and {Douspis}, M. and {Ducout}, A. and {Dupac}, X. and {Dusini}, S. and {Efstathiou}, G. and {Elsner}, F. and {En{\ss}lin}, T.~A. and {Eriksen}, H.~K. and {Fantaye}, Y. and {Farhang}, M. and {Fergusson}, J. and {Fernandez-Cobos}, R. and {Finelli}, F. and {Forastieri}, F. and {Frailis}, M. and {Fraisse}, A.~A. and {Franceschi}, E. and {Frolov}, A. and {Galeotta}, S. and {Galli}, S. and {Ganga}, K. and {G{\'e}nova-Santos}, R.~T. and {Gerbino}, M. and {Ghosh}, T. and {Gonz{\'a}lez-Nuevo}, J. and {G{\'o}rski}, K.~M. and {Gratton}, S. and {Gruppuso}, A. and {Gudmundsson}, J.~E. and {Hamann}, J. and {Handley}, W. and {Hansen}, F.~K. and {Herranz}, D. and {Hildebrandt}, S.~R. and {Hivon}, E. and {Huang}, Z. and {Jaffe}, A.~H. and {Jones}, W.~C. and {Karakci}, A. and {Keih{\"a}nen}, E. and {Keskitalo}, R. and {Kiiveri}, K. and {Kim}, J. and {Kisner}, T.~S. and {Knox}, L. and {Krachmalnicoff}, N. and {Kunz}, M. and {Kurki-Suonio}, H. and {Lagache}, G. and {Lamarre}, J. -M. and {Lasenby}, A. and {Lattanzi}, M. and {Lawrence}, C.~R. and {Le Jeune}, M. and {Lemos}, P. and {Lesgourgues}, J. and {Levrier}, F. and {Lewis}, A. and {Liguori}, M. and {Lilje}, P.~B. and {Lilley}, M. and {Lindholm}, V. and {L{\'o}pez-Caniego}, M. and {Lubin}, P.~M. and {Ma}, Y. -Z. and {Mac{\'\i}as-P{\'e}rez}, J.~F. and {Maggio}, G. and {Maino}, D. and {Mandolesi}, N. and {Mangilli}, A. and {Marcos-Caballero}, A. and {Maris}, M. and {Martin}, P.~G. and {Martinelli}, M. and {Mart{\'\i}nez-Gonz{\'a}lez}, E. and {Matarrese}, S. and {Mauri}, N. and {McEwen}, J.~D. and {Meinhold}, P.~R. and {Melchiorri}, A. and {Mennella}, A. and {Migliaccio}, M. and {Millea}, M. and {Mitra}, S. and {Miville-Desch{\^e}nes}, M. -A. and {Molinari}, D. and {Montier}, L. and {Morgante}, G. and {Moss}, A. and {Natoli}, P. and {N{\o}rgaard-Nielsen}, H.~U. and {Pagano}, L. and {Paoletti}, D. and {Partridge}, B. and {Patanchon}, G. and {Peiris}, H.~V. and {Perrotta}, F. and {Pettorino}, V. and {Piacentini}, F. and {Polastri}, L. and {Polenta}, G. and {Puget}, J. -L. and {Rachen}, J.~P. and {Reinecke}, M. and {Remazeilles}, M. and {Renzi}, A. and {Rocha}, G. and {Rosset}, C. and {Roudier}, G. and {Rubi{\~n}o-Mart{\'\i}n}, J.~A. and {Ruiz-Granados}, B. and {Salvati}, L. and {Sandri}, M. and {Savelainen}, M. and {Scott}, D. and {Shellard}, E.~P.~S. and {Sirignano}, C. and {Sirri}, G. and {Spencer}, L.~D. and {Sunyaev}, R. and {Suur-Uski}, A. -S. and {Tauber}, J.~A. and {Tavagnacco}, D. and {Tenti}, M. and {Toffolatti}, L. and {Tomasi}, M. and {Trombetti}, T. and {Valenziano}, L. and {Valiviita}, J. and {Van Tent}, B. and {Vibert}, L. and {Vielva}, P. and {Villa}, F. and {Vittorio}, N. and {Wandelt}, B.~D. and {Wehus}, I.~K. and {White}, M. and {White}, S.~D.~M. and {Zacchei}, A. and {Zonca}, A.},
        title = "{Planck 2018 results. VI. Cosmological parameters}",
      journal = {\aap},
         year = 2020,
        month = sep,
       volume = {641},
          eid = {A6},
        pages = {A6},
          doi = {10.1051/0004-6361/201833910},
archivePrefix = {arXiv},
       eprint = {1807.06209},
 primaryClass = {astro-ph.CO},
       adsurl = {https://ui.adsabs.harvard.edu/abs/2020A&A...641A...6P}
}

@ARTICLE{Bonaca2017,
       author = {{Bonaca}, Ana and {Conroy}, Charlie and {Wetzel}, Andrew and {Hopkins}, Philip F. and {Kere{\v{s}}}, Du{\v{s}}an},
        title = "{Gaia Reveals a Metal-rich, in situ Component of the Local Stellar Halo}",
      journal = {\apj},
         year = 2017,
        month = aug,
       volume = {845},
       number = {2},
          eid = {101},
        pages = {101},
          doi = {10.3847/1538-4357/aa7d0c},
archivePrefix = {arXiv},
       eprint = {1704.05463},
 primaryClass = {astro-ph.GA},
       adsurl = {https://ui.adsabs.harvard.edu/abs/2017ApJ...845..101B}
}

@ARTICLE{Ma2017,
       author = {{Ma}, Xiangcheng and {Hopkins}, Philip F. and {Wetzel}, Andrew R. and {Kirby}, Evan N. and {Angl{\'e}s-Alc{\'a}zar}, Daniel and {Faucher-Gigu{\`e}re}, Claude-Andr{\'e} and {Kere{\v{s}}}, Du{\v{s}}an and {Quataert}, Eliot},
        title = "{The structure and dynamical evolution of the stellar disc of a simulated Milky Way-mass galaxy}",
      journal = {\mnras},
         year = 2017,
        month = may,
       volume = {467},
       number = {2},
        pages = {2430-2444},
          doi = {10.1093/mnras/stx273},
archivePrefix = {arXiv},
       eprint = {1608.04133},
 primaryClass = {astro-ph.GA},
       adsurl = {https://ui.adsabs.harvard.edu/abs/2017MNRAS.467.2430M}
}

@ARTICLE{Sanderson2020,
       author = {{Sanderson}, Robyn E. and {Wetzel}, Andrew and {Loebman}, Sarah and {Sharma}, Sanjib and {Hopkins}, Philip F. and {Garrison-Kimmel}, Shea and {Faucher-Gigu{\`e}re}, Claude-Andr{\'e} and {Kere{\v{s}}}, Du{\v{s}}an and {Quataert}, Eliot},
        title = "{Synthetic Gaia Surveys from the FIRE Cosmological Simulations of Milky Way-mass Galaxies}",
      journal = {\apjs},
         year = 2020,
        month = jan,
       volume = {246},
       number = {1},
          eid = {6},
        pages = {6},
          doi = {10.3847/1538-4365/ab5b9d},
archivePrefix = {arXiv},
       eprint = {1806.10564},
 primaryClass = {astro-ph.GA},
       adsurl = {https://ui.adsabs.harvard.edu/abs/2020ApJS..246....6S}
}

@ARTICLE{Bellardini2021,
       author = {{Bellardini}, Matthew A. and {Wetzel}, Andrew and {Loebman}, Sarah R. and {Faucher-Gigu{\`e}re}, Claude-Andr{\'e} and {Ma}, Xiangcheng and {Feldmann}, Robert},
        title = "{3D gas-phase elemental abundances across the formation histories of Milky Way-mass galaxies in the FIRE simulations: initial conditions for chemical tagging}",
      journal = {\mnras},
         year = 2021,
        month = aug,
       volume = {505},
       number = {3},
        pages = {4586-4607},
          doi = {10.1093/mnras/stab1606},
archivePrefix = {arXiv},
       eprint = {2102.06220},
 primaryClass = {astro-ph.GA},
       adsurl = {https://ui.adsabs.harvard.edu/abs/2021MNRAS.505.4586B}
}

@ARTICLE{Samuel2020,
       author = {{Samuel}, Jenna and {Wetzel}, Andrew and {Tollerud}, Erik and {Garrison-Kimmel}, Shea and {Loebman}, Sarah and {El-Badry}, Kareem and {Hopkins}, Philip F. and {Boylan-Kolchin}, Michael and {Faucher-Gigu{\`e}re}, Claude-Andr{\'e} and {Bullock}, James S. and {Benincasa}, Samantha and {Bailin}, Jeremy},
        title = "{A profile in FIRE: resolving the radial distributions of satellite galaxies in the Local Group with simulations}",
      journal = {\mnras},
         year = 2020,
        month = jan,
       volume = {491},
       number = {1},
        pages = {1471-1490},
          doi = {10.1093/mnras/stz3054},
archivePrefix = {arXiv},
       eprint = {1904.11508},
 primaryClass = {astro-ph.GA},
       adsurl = {https://ui.adsabs.harvard.edu/abs/2020MNRAS.491.1471S}
}

@ARTICLE{Garrison-Kimmel2019b,
       author = {{Garrison-Kimmel}, Shea and {Wetzel}, Andrew and {Hopkins}, Philip F. and {Sanderson}, Robyn and {El-Badry}, Kareem and {Graus}, Andrew and {Chan}, T.~K. and {Feldmann}, Robert and {Boylan-Kolchin}, Michael and {Hayward}, Christopher C. and {Bullock}, James S. and {Fitts}, Alex and {Samuel}, Jenna and {Wheeler}, Coral and {Kere{\v{s}}}, Du{\v{s}}an and {Faucher-Gigu{\`e}re}, Claude-Andr{\'e}},
        title = "{Star formation histories of dwarf galaxies in the FIRE simulations: dependence on mass and Local Group environment}",
      journal = {\mnras},
         year = 2019,
        month = nov,
       volume = {489},
       number = {4},
        pages = {4574-4588},
          doi = {10.1093/mnras/stz2507},
archivePrefix = {arXiv},
       eprint = {1903.10515},
 primaryClass = {astro-ph.GA},
       adsurl = {https://ui.adsabs.harvard.edu/abs/2019MNRAS.489.4574G}
}

@ARTICLE{Panithanpaisal2021,
       author = {{Panithanpaisal}, Nondh and {Sanderson}, Robyn E. and {Wetzel}, Andrew and {Cunningham}, Emily C. and {Bailin}, Jeremy and {Faucher-Gigu{\`e}re}, Claude-Andr{\'e}},
        title = "{The Galaxy Progenitors of Stellar Streams around Milky Way-mass Galaxies in the FIRE Cosmological Simulations}",
      journal = {\apj},
         year = 2021,
        month = oct,
       volume = {920},
       number = {1},
          eid = {10},
        pages = {10},
          doi = {10.3847/1538-4357/ac1109},
archivePrefix = {arXiv},
       eprint = {2104.09660},
 primaryClass = {astro-ph.GA},
       adsurl = {https://ui.adsabs.harvard.edu/abs/2021ApJ...920...10P}
}

@ARTICLE{Cunningham2021,
       author = {{Cunningham}, Emily C. and {Sanderson}, Robyn E. and {Johnston}, Kathryn V. and {Panithanpaisal}, Nondh and {Ness}, Melissa K. and {Wetzel}, Andrew and {Loebman}, Sarah R. and {Escala}, Ivanna and {Horta}, Danny and {Faucher-Gigu{\`e}re}, Claude-Andr{\'e}},
        title = "{Reading the CARDs: The Imprint of Accretion History in the Chemical Abundances of the Milky Way's Stellar Halo}",
      journal = {\apj},
         year = 2022,
        month = aug,
       volume = {934},
       number = {2},
          eid = {172},
        pages = {172},
          doi = {10.3847/1538-4357/ac78ea},
archivePrefix = {arXiv},
       eprint = {2110.02957},
 primaryClass = {astro-ph.GA},
       adsurl = {https://ui.adsabs.harvard.edu/abs/2022ApJ...934..172C}
}

@ARTICLE{Shipp2023,
       author = {{Shipp}, Nora and {Panithanpaisal}, Nondh and {Necib}, Lina and {Sanderson}, Robyn and {Erkal}, Denis and {Li}, Ting S. and {Santistevan}, Isaiah B. and {Wetzel}, Andrew and {Cullinane}, Lara R. and {Ji}, Alexander P. and {Koposov}, Sergey E. and {Kuehn}, Kyler and {Lewis}, Geraint F. and {Pace}, Andrew B. and {Zucker}, Daniel B. and {Bland-Hawthorn}, Joss and {Cunningham}, Emily C. and {Kim}, Stacy Y. and {Lilleengen}, Sophia and {Moreno}, Jorge and {Sharma}, Sanjib and {S Collaboration} and {FIRE Collaboration}},
        title = "{Streams on FIRE: Populations of Detectable Stellar Streams in the Milky Way and FIRE}",
      journal = {\apj},
         year = 2023,
        month = jun,
       volume = {949},
       number = {2},
          eid = {44},
        pages = {44},
          doi = {10.3847/1538-4357/acc582},
archivePrefix = {arXiv},
       eprint = {2208.02255},
 primaryClass = {astro-ph.GA},
       adsurl = {https://ui.adsabs.harvard.edu/abs/2023ApJ...949...44S}
}

@ARTICLE{Belokurov2006,
       author = {{Belokurov}, V. and {Zucker}, D.~B. and {Evans}, N.~W. and {Gilmore}, G. and {Vidrih}, S. and {Bramich}, D.~M. and {Newberg}, H.~J. and {Wyse}, R.~F.~G. and {Irwin}, M.~J. and {Fellhauer}, M. and {Hewett}, P.~C. and {Walton}, N.~A. and {Wilkinson}, M.~I. and {Cole}, N. and {Yanny}, B. and {Rockosi}, C.~M. and {Beers}, T.~C. and {Bell}, E.~F. and {Brinkmann}, J. and {Ivezi{\'c}}, {\v{Z}}. and {Lupton}, R.},
        title = "{The Field of Streams: Sagittarius and Its Siblings}",
      journal = {\apjl},
         year = 2006,
        month = may,
       volume = {642},
       number = {2},
        pages = {L137-L140},
          doi = {10.1086/504797},
archivePrefix = {arXiv},
       eprint = {astro-ph/0605025},
 primaryClass = {astro-ph},
       adsurl = {https://ui.adsabs.harvard.edu/abs/2006ApJ...642L.137B}
}

@ARTICLE{Monachesi2019,
       author = {{Monachesi}, Antonela and {G{\'o}mez}, Facundo A. and {Grand}, Robert J.~J. and {Simpson}, Christine M. and {Kauffmann}, Guinevere and {Bustamante}, Sebasti{\'a}n and {Marinacci}, Federico and {Pakmor}, R{\"u}diger and {Springel}, Volker and {Frenk}, Carlos S. and {White}, Simon D.~M. and {Tissera}, Patricia B.},
        title = "{The Auriga stellar haloes: connecting stellar population properties with accretion and merging history}",
      journal = {\mnras},
         year = 2019,
        month = may,
       volume = {485},
       number = {2},
        pages = {2589-2616},
          doi = {10.1093/mnras/stz538},
archivePrefix = {arXiv},
       eprint = {1804.07798},
 primaryClass = {astro-ph.GA},
       adsurl = {https://ui.adsabs.harvard.edu/abs/2019MNRAS.485.2589M}
}

@ARTICLE{Behroozi2013a,
       author = {{Behroozi}, Peter S. and {Wechsler}, Risa H. and {Wu}, Hao-Yi},
        title = "{The ROCKSTAR Phase-space Temporal Halo Finder and the Velocity Offsets of Cluster Cores}",
      journal = {\apj},
         year = 2013,
        month = jan,
       volume = {762},
       number = {2},
          eid = {109},
        pages = {109},
          doi = {10.1088/0004-637X/762/2/109},
archivePrefix = {arXiv},
       eprint = {1110.4372},
 primaryClass = {astro-ph.CO},
       adsurl = {https://ui.adsabs.harvard.edu/abs/2013ApJ...762..109B}
}

@ARTICLE{Behroozi2013b, author = {{Behroozi}, Peter S. and {Wechsler}, Risa H. and {Wu}, Hao-Yi and {Busha}, Michael T. and {Klypin}, Anatoly A. and {Primack}, Joel R.},
        title = "{Gravitationally Consistent Halo Catalogs and Merger Trees for Precision Cosmology}",
      journal = {\apj},
         year = 2013,
        month = jan,
       volume = {763},
       number = {1},
          eid = {18},
        pages = {18},
          doi = {10.1088/0004-637X/763/1/18},
archivePrefix = {arXiv},
       eprint = {1110.4370},
 primaryClass = {astro-ph.CO},
       adsurl = {https://ui.adsabs.harvard.edu/abs/2013ApJ...763...18B}
}

@ARTICLE{Horta2024,
       author = {{Horta}, Danny and {Cunningham}, Emily C. and {Sanderson}, Robyn and {Johnston}, Kathryn V. and {Deason}, Alis and {Wetzel}, Andrew and {McCluskey}, Fiona and {Garavito-Camargo}, Nicol{\'a}s and {Necib}, Lina and {Faucher-Gigu{\`e}re}, Claude-Andr{\'e} and {Arora}, Arpit and {Gandhi}, Pratik J.},
        title = "{The proto-galaxy of Milky Way-mass haloes in the FIRE simulations}",
      journal = {\mnras},
         year = 2024,
        month = feb,
       volume = {527},
       number = {4},
        pages = {9810-9825},
          doi = {10.1093/mnras/stad3834},
archivePrefix = {arXiv},
       eprint = {2307.15741},
 primaryClass = {astro-ph.GA},
       adsurl = {https://ui.adsabs.harvard.edu/abs/2024MNRAS.527.9810H}
}

@ARTICLE{Leitherer1999,
       author = {{Leitherer}, Claus and {Schaerer}, Daniel and {Goldader}, Jeffrey D. and {Delgado}, Rosa M. Gonz{\'a}lez and {Robert}, Carmelle and {Kune}, Denis Foo and {de Mello}, Du{\'\i}lia F. and {Devost}, Daniel and {Heckman}, Timothy M.},
        title = "{Starburst99: Synthesis Models for Galaxies with Active Star Formation}",
      journal = {\apjs},
         year = 1999,
        month = jul,
       volume = {123},
       number = {1},
        pages = {3-40},
          doi = {10.1086/313233},
archivePrefix = {arXiv},
       eprint = {astro-ph/9902334},
 primaryClass = {astro-ph},
       adsurl = {https://ui.adsabs.harvard.edu/abs/1999ApJS..123....3L}
}

@ARTICLE{Planck2014,
       author = {{Planck Collaboration} and {Ade}, P.~A.~R. and {Aghanim}, N. and {Armitage-Caplan}, C. and {Arnaud}, M. and {Ashdown}, M. and {Atrio-Barandela}, F. and {Aumont}, J. and {Baccigalupi}, C. and {Banday}, A.~J. and {Barreiro}, R.~B. and {Bartlett}, J.~G. and {Battaner}, E. and {Benabed}, K. and {Beno{\^\i}t}, A. and {Benoit-L{\'e}vy}, A. and {Bernard}, J. -P. and {Bersanelli}, M. and {Bielewicz}, P. and {Bobin}, J. and {Bock}, J.~J. and {Bonaldi}, A. and {Bond}, J.~R. and {Borrill}, J. and {Bouchet}, F.~R. and {Bridges}, M. and {Bucher}, M. and {Burigana}, C. and {Butler}, R.~C. and {Calabrese}, E. and {Cappellini}, B. and {Cardoso}, J. -F. and {Catalano}, A. and {Challinor}, A. and {Chamballu}, A. and {Chary}, R. -R. and {Chen}, X. and {Chiang}, H.~C. and {Chiang}, L. -Y. and {Christensen}, P.~R. and {Church}, S. and {Clements}, D.~L. and {Colombi}, S. and {Colombo}, L.~P.~L. and {Couchot}, F. and {Coulais}, A. and {Crill}, B.~P. and {Curto}, A. and {Cuttaia}, F. and {Danese}, L. and {Davies}, R.~D. and {Davis}, R.~J. and {de Bernardis}, P. and {de Rosa}, A. and {de Zotti}, G. and {Delabrouille}, J. and {Delouis}, J. -M. and {D{\'e}sert}, F. -X. and {Dickinson}, C. and {Diego}, J.~M. and {Dolag}, K. and {Dole}, H. and {Donzelli}, S. and {Dor{\'e}}, O. and {Douspis}, M. and {Dunkley}, J. and {Dupac}, X. and {Efstathiou}, G. and {Elsner}, F. and {En{\ss}lin}, T.~A. and {Eriksen}, H.~K. and {Finelli}, F. and {Forni}, O. and {Frailis}, M. and {Fraisse}, A.~A. and {Franceschi}, E. and {Gaier}, T.~C. and {Galeotta}, S. and {Galli}, S. and {Ganga}, K. and {Giard}, M. and {Giardino}, G. and {Giraud-H{\'e}raud}, Y. and {Gjerl{\o}w}, E. and {Gonz{\'a}lez-Nuevo}, J. and {G{\'o}rski}, K.~M. and {Gratton}, S. and {Gregorio}, A. and {Gruppuso}, A. and {Gudmundsson}, J.~E. and {Haissinski}, J. and {Hamann}, J. and {Hansen}, F.~K. and {Hanson}, D. and {Harrison}, D. and {Henrot-Versill{\'e}}, S. and {Hern{\'a}ndez-Monteagudo}, C. and {Herranz}, D. and {Hildebrandt}, S.~R. and {Hivon}, E. and {Hobson}, M. and {Holmes}, W.~A. and {Hornstrup}, A. and {Hou}, Z. and {Hovest}, W. and {Huffenberger}, K.~M. and {Jaffe}, A.~H. and {Jaffe}, T.~R. and {Jewell}, J. and {Jones}, W.~C. and {Juvela}, M. and {Keih{\"a}nen}, E. and {Keskitalo}, R. and {Kisner}, T.~S. and {Kneissl}, R. and {Knoche}, J. and {Knox}, L. and {Kunz}, M. and {Kurki-Suonio}, H. and {Lagache}, G. and {L{\"a}hteenm{\"a}ki}, A. and {Lamarre}, J. -M. and {Lasenby}, A. and {Lattanzi}, M. and {Laureijs}, R.~J. and {Lawrence}, C.~R. and {Leach}, S. and {Leahy}, J.~P. and {Leonardi}, R. and {Le{\'o}n-Tavares}, J. and {Lesgourgues}, J. and {Lewis}, A. and {Liguori}, M. and {Lilje}, P.~B. and {Linden-V{\o}rnle}, M. and {L{\'o}pez-Caniego}, M. and {Lubin}, P.~M. and {Mac{\'\i}as-P{\'e}rez}, J.~F. and {Maffei}, B. and {Maino}, D. and {Mandolesi}, N. and {Maris}, M. and {Marshall}, D.~J. and {Martin}, P.~G. and {Mart{\'\i}nez-Gonz{\'a}lez}, E. and {Masi}, S. and {Massardi}, M. and {Matarrese}, S. and {Matthai}, F. and {Mazzotta}, P. and {Meinhold}, P.~R. and {Melchiorri}, A. and {Melin}, J. -B. and {Mendes}, L. and {Menegoni}, E. and {Mennella}, A. and {Migliaccio}, M. and {Millea}, M. and {Mitra}, S. and {Miville-Desch{\^e}nes}, M. -A. and {Moneti}, A. and {Montier}, L. and {Morgante}, G. and {Mortlock}, D. and {Moss}, A. and {Munshi}, D. and {Murphy}, J.~A. and {Naselsky}, P. and {Nati}, F. and {Natoli}, P. and {Netterfield}, C.~B. and {N{\o}rgaard-Nielsen}, H.~U. and {Noviello}, F. and {Novikov}, D. and {Novikov}, I. and {O'Dwyer}, I.~J. and {Osborne}, S. and {Oxborrow}, C.~A. and {Paci}, F. and {Pagano}, L. and {Pajot}, F. and {Paladini}, R. and {Paoletti}, D. and {Partridge}, B. and {Pasian}, F. and {Patanchon}, G. and {Pearson}, D. and {Pearson}, T.~J. and {Peiris}, H.~V. and {Perdereau}, O. and {Perotto}, L. and {Perrotta}, F. and {Pettorino}, V. and {Piacentini}, F. and {Piat}, M. and {Pierpaoli}, E. and {Pietrobon}, D. and {Plaszczynski}, S. and {Platania}, P. and {Pointecouteau}, E. and {Polenta}, G. and {Ponthieu}, N. and {Popa}, L. and {Poutanen}, T. and {Pratt}, G.~W. and {Pr{\'e}zeau}, G. and {Prunet}, S. and {Puget}, J. -L. and {Rachen}, J.~P. and {Reach}, W.~T. and {Rebolo}, R. and {Reinecke}, M. and {Remazeilles}, M. and {Renault}, C. and {Ricciardi}, S. and {Riller}, T. and {Ristorcelli}, I. and {Rocha}, G. and {Rosset}, C. and {Roudier}, G. and {Rowan-Robinson}, M. and {Rubi{\~n}o-Mart{\'\i}n}, J.~A. and {Rusholme}, B. and {Sandri}, M. and {Santos}, D. and {Savelainen}, M. and {Savini}, G. and {Scott}, D. and {Seiffert}, M.~D. and {Shellard}, E.~P.~S. and {Spencer}, L.~D. and {Starck}, J. -L. and {Stolyarov}, V. and {Stompor}, R. and {Sudiwala}, R. and {Sunyaev}, R. and {Sureau}, F. and {Sutton}, D. and {Suur-Uski}, A. -S. and {Sygnet}, J. -F. and {Tauber}, J.~A. and {Tavagnacco}, D. and {Terenzi}, L. and {Toffolatti}, L. and {Tomasi}, M. and {Tristram}, M. and {Tucci}, M. and {Tuovinen}, J. and {T{\"u}rler}, M. and {Umana}, G. and {Valenziano}, L. and {Valiviita}, J. and {Van Tent}, B. and {Vielva}, P. and {Villa}, F. and {Vittorio}, N. and {Wade}, L.~A. and {Wandelt}, B.~D. and {Wehus}, I.~K. and {White}, M. and {White}, S.~D.~M. and {Wilkinson}, A. and {Yvon}, D. and {Zacchei}, A. and {Zonca}, A.},
        title = "{Planck 2013 results. XVI. Cosmological parameters}",
      journal = {\aap},
         year = 2014,
        month = nov,
       volume = {571},
          eid = {A16},
        pages = {A16},
          doi = {10.1051/0004-6361/201321591},
archivePrefix = {arXiv},
       eprint = {1303.5076},
 primaryClass = {astro-ph.CO},
       adsurl = {https://ui.adsabs.harvard.edu/abs/2014A&A...571A..16P}
}

@ARTICLE{Pontzen2014,
       author = {{Pontzen}, Andrew and {Governato}, Fabio},
        title = "{Cold dark matter heats up}",
      journal = {\nat},
         year = 2014,
        month = feb,
       volume = {506},
       number = {7487},
        pages = {171-178},
          doi = {10.1038/nature12953},
archivePrefix = {arXiv},
       eprint = {1402.1764},
 primaryClass = {astro-ph.CO},
       adsurl = {https://ui.adsabs.harvard.edu/abs/2014Natur.506..171P}
}

@ARTICLE{Lazar2020,
       author = {{Lazar}, Alexandres and {Bullock}, James S. and {Boylan-Kolchin}, Michael and {Chan}, T.~K. and {Hopkins}, Philip F. and {Graus}, Andrew S. and {Wetzel}, Andrew and {El-Badry}, Kareem and {Wheeler}, Coral and {Straight}, Maria C. and {Kere{\v{s}}}, Du{\v{s}}an and {Faucher-Gigu{\`e}re}, Claude-Andr{\'e} and {Fitts}, Alex and {Garrison-Kimmel}, Shea},
        title = "{A dark matter profile to model diverse feedback-induced core sizes of {\ensuremath{\Lambda}}CDM haloes}",
      journal = {\mnras},
         year = 2020,
        month = sep,
       volume = {497},
       number = {2},
        pages = {2393-2417},
          doi = {10.1093/mnras/staa2101},
archivePrefix = {arXiv},
       eprint = {2004.10817},
 primaryClass = {astro-ph.GA},
       adsurl = {https://ui.adsabs.harvard.edu/abs/2020MNRAS.497.2393L}
}

@ARTICLE{Wetzel2023,
       author = {{Wetzel}, Andrew and {Hayward}, Christopher C. and {Sanderson}, Robyn E. and {Ma}, Xiangcheng and {Angl{\'e}s-Alc{\'a}zar}, Daniel and {Feldmann}, Robert and {Chan}, T.~K. and {El-Badry}, Kareem and {Wheeler}, Coral and {Garrison-Kimmel}, Shea and {Nikakhtar}, Farnik and {Panithanpaisal}, Nondh and {Arora}, Arpit and {Gurvich}, Alexander B. and {Samuel}, Jenna and {Sameie}, Omid and {Pandya}, Viraj and {Hafen}, Zachary and {Hummels}, Cameron and {Loebman}, Sarah and {Boylan-Kolchin}, Michael and {Bullock}, James S. and {Faucher-Gigu{\`e}re}, Claude-Andr{\'e} and {Kere{\v{s}}}, Du{\v{s}}an and {Quataert}, Eliot and {Hopkins}, Philip F.},
        title = "{Public Data Release of the FIRE-2 Cosmological Zoom-in Simulations of Galaxy Formation}",
      journal = {\apjs},
         year = 2023,
        month = apr,
       volume = {265},
       number = {2},
          eid = {44},
        pages = {44},
          doi = {10.3847/1538-4365/acb99a},
archivePrefix = {arXiv},
       eprint = {2202.06969},
 primaryClass = {astro-ph.GA},
       adsurl = {https://ui.adsabs.harvard.edu/abs/2023ApJS..265...44W}
}

@software{GizmoAnalysis,
       author = {{Wetzel}, Andrew and {Garrison-Kimmel}, Shea},
        title = "{GizmoAnalysis: Read and analyze Gizmo simulations}",
 howpublished = {Astrophysics Source Code Library, record ascl:2002.015},
         year = 2020,
        month = feb,
          eid = {ascl:2002.015},
       adsurl = {https://ui.adsabs.harvard.edu/abs/2020ascl.soft02015W}
}

@ARTICLE{Yu2023,
       author = {{Yu}, Sijie and {Bullock}, James S. and {Gurvich}, Alexander B. and {Hafen}, Zachary and {Stern}, Jonathan and {Boylan-Kolchin}, Michael and {Faucher-Gigu{\`e}re}, Claude-Andr{\'e} and {Wetzel}, Andrew and {Hopkins}, Philip F. and {Moreno}, Jorge},
        title = "{Born this way: thin disc, thick disc, and isotropic spheroid formation in FIRE-2 Milky Way-mass galaxy simulations}",
      journal = {\mnras},
         year = 2023,
        month = aug,
       volume = {523},
       number = {4},
        pages = {6220-6238},
          doi = {10.1093/mnras/stad1806},
archivePrefix = {arXiv},
       eprint = {2210.03845},
 primaryClass = {astro-ph.GA},
       adsurl = {https://ui.adsabs.harvard.edu/abs/2023MNRAS.523.6220Y}
}

@ARTICLE{McCluskey2024,
       author = {{McCluskey}, Fiona and {Wetzel}, Andrew and {Loebman}, Sarah R. and {Moreno}, Jorge and {Faucher-Gigu{\`e}re}, Claude-Andr{\'e} and {Hopkins}, Philip F.},
        title = "{Disc settling and dynamical heating: histories of Milky Way-mass stellar discs across cosmic time in the FIRE simulations}",
      journal = {\mnras},
         year = 2024,
        month = jan,
       volume = {527},
       number = {3},
        pages = {6926-6949},
          doi = {10.1093/mnras/stad3547},
archivePrefix = {arXiv},
       eprint = {2303.14210},
 primaryClass = {astro-ph.GA},
       adsurl = {https://ui.adsabs.harvard.edu/abs/2024MNRAS.527.6926M}
}

@ARTICLE{Chandra2025,
       author = {{Chandra}, Vedant and {Cargile}, Phillip A. and {Ji}, Alexander P. and {Conroy}, Charlie and {Rix}, Hans-Walter and {Cunningham}, Emily and {Dias}, Bruno and {Laporte}, Chervin and {Cerny}, William and {Limberg}, Guilherme and {Bandyopadhyay}, Avrajit and {Bonaca}, Ana and {Casey}, Andrew R. and {Donor}, John and {Fernandez-Trincado}, Jose G. and {Frinchaboy}, Peter M. and {Gupta}, Pramod and {Hawkins}, Keith and {Johnson}, Jennifer A. and {Kollmeier}, Juna A. and {Lucey}, Madeline and {Medan}, Ilija and {Meszaros}, Szabolcs and {Morrison}, Sean and {Sanchez-Gallego}, Jose and {Saydjari}, Andrew K. and {Sayres}, Conor and {Schlaufman}, Kevin C. and {Stassun}, Keivan G. and {Tayar}, Jamie and {Way}, Zachary},
        title = "{Mapping the Distant and Metal-Poor Milky Way with SDSS-V}",
      journal = {arXiv e-prints},
         year = 2025,
        month = aug,
          eid = {arXiv:2508.00978},
        pages = {arXiv:2508.00978},
archivePrefix = {arXiv},
       eprint = {2508.00978},
 primaryClass = {astro-ph.GA},
       adsurl = {https://ui.adsabs.harvard.edu/abs/2025arXiv250800978C}
}

@ARTICLE{Rix2022,
       author = {{Rix}, Hans-Walter and {Chandra}, Vedant and {Andrae}, Ren{\'e} and {Price-Whelan}, Adrian M. and {Weinberg}, David H. and {Conroy}, Charlie and {Fouesneau}, Morgan and {Hogg}, David W. and {De Angeli}, Francesca and {Naidu}, Rohan P. and {Xiang}, Maosheng and {Ruz-Mieres}, Daniela},
        title = "{The Poor Old Heart of the Milky Way}",
      journal = {\apj},
         year = 2022,
        month = dec,
       volume = {941},
       number = {1},
          eid = {45},
        pages = {45},
          doi = {10.3847/1538-4357/ac9e01},
archivePrefix = {arXiv},
       eprint = {2209.02722},
 primaryClass = {astro-ph.GA},
       adsurl = {https://ui.adsabs.harvard.edu/abs/2022ApJ...941...45R}
}

@ARTICLE{Hopkins2018,
       author = {{Hopkins}, Philip F. and {Wetzel}, Andrew and {Kere{\v{s}}}, Du{\v{s}}an and {Faucher-Gigu{\`e}re}, Claude-Andr{\'e} and {Quataert}, Eliot and {Boylan-Kolchin}, Michael and {Murray}, Norman and {Hayward}, Christopher C. and {El-Badry}, Kareem},
        title = "{How to model supernovae in simulations of star and galaxy formation}",
      journal = {\mnras},
         year = 2018,
        month = jun,
       volume = {477},
       number = {2},
        pages = {1578-1603},
          doi = {10.1093/mnras/sty674},
archivePrefix = {arXiv},
       eprint = {1707.07010},
 primaryClass = {astro-ph.GA},
       adsurl = {https://ui.adsabs.harvard.edu/abs/2018MNRAS.477.1578H}
}

@article{Bullock2001,
   title={Hierarchical Galaxy Formation and Substructure in the Galaxy’s Stellar Halo},
   volume={548},
   ISSN={1538-4357},
   url={http://dx.doi.org/10.1086/318681},
   DOI={10.1086/318681},
   number={1},
   journal={The Astrophysical Journal},
   publisher={American Astronomical Society},
   author={Bullock, James S. and Kravtsov, Andrey V. and Weinberg, David H.},
   year={2001},
   month=feb, pages={33–46} }

@article{Mackereth2018,
   title={The origin of accreted stellar halo populations in the Milky Way using APOGEE,Gaia, and the EAGLE simulations},
   volume={482},
   ISSN={1365-2966},
   url={http://dx.doi.org/10.1093/mnras/sty2955},
   DOI={10.1093/mnras/sty2955},
   number={3},
   journal={Monthly Notices of the Royal Astronomical Society},
   publisher={Oxford University Press (OUP)},
   author={Mackereth, J Ted and Schiavon, Ricardo P and Pfeffer, Joel and Hayes, Christian R and Bovy, Jo and Anguiano, Borja and Prieto, Carlos Allende and Hasselquist, Sten and Holtzman, Jon and Johnson, Jennifer A and Majewski, Steven R and O’Connell, Robert and Shetrone, Matthew and Tissera, Patricia B and Fernández-Trincado, J G},
   year={2018},
   month=nov, pages={3426–3442} }

@article{Kirby2008b,
   title={Uncovering Extremely Metal-Poor Stars in the Milky Way’s Ultrafaint Dwarf Spheroidal Satellite Galaxies},
   volume={685},
   ISSN={1538-4357},
   url={http://dx.doi.org/10.1086/592432},
   DOI={10.1086/592432},
   number={1},
   journal={The Astrophysical Journal},
   publisher={American Astronomical Society},
   author={Kirby, Evan N. and Simon, Joshua D. and Geha, Marla and Guhathakurta, Puragra and Frebel, Anna},
   year={2008},
   month=sep, pages={L43–L46} }

@article{Bovill2011,
   title={WHERE ARE THE FOSSILS OF THE FIRST GALAXIES? I. LOCAL VOLUME MAPS AND PROPERTIES OF THE UNDETECTED DWARFS},
   volume={741},
   ISSN={1538-4357},
   url={http://dx.doi.org/10.1088/0004-637X/741/1/17},
   DOI={10.1088/0004-637x/741/1/17},
   number={1},
   journal={The Astrophysical Journal},
   publisher={American Astronomical Society},
   author={Bovill, Mia S. and Ricotti, Massimo},
   year={2011},
   month=oct, pages={17} }

@article{Beniamini2018,
   title={Retainment of r-process material in dwarf galaxies},
   volume={478},
   ISSN={1365-2966},
   url={http://dx.doi.org/10.1093/mnras/sty1035},
   DOI={10.1093/mnras/sty1035},
   number={2},
   journal={Monthly Notices of the Royal Astronomical Society},
   publisher={Oxford University Press (OUP)},
   author={Beniamini, Paz and Dvorkin, Irina and Silk, Joe},
   year={2018},
   month=apr, pages={1994–2005} }

@ARTICLE{Aumer2013b,
       author = {{Aumer}, Michael and {White}, Simon D.~M. and {Naab}, Thorsten and {Scannapieco}, Cecilia},
        title = "{Towards a more realistic population of bright spiral galaxies in cosmological simulations}",
      journal = {\mnras},
         year = 2013,
        month = oct,
       volume = {434},
       number = {4},
        pages = {3142-3164},
          doi = {10.1093/mnras/stt1230},
archivePrefix = {arXiv},
       eprint = {1304.1559},
 primaryClass = {astro-ph.GA},
       adsurl = {https://ui.adsabs.harvard.edu/abs/2013MNRAS.434.3142A}
}

@ARTICLE{Aumer2013a,
       author = {{Aumer}, Michael and {White}, Simon D.~M.},
        title = "{Idealized models for galactic disc formation and evolution in `realistic' {\ensuremath{\Lambda}}CDM haloes}",
      journal = {\mnras},
         year = 2013,
        month = jan,
       volume = {428},
       number = {2},
        pages = {1055-1076},
          doi = {10.1093/mnras/sts083},
archivePrefix = {arXiv},
       eprint = {1203.1190},
 primaryClass = {astro-ph.GA},
       adsurl = {https://ui.adsabs.harvard.edu/abs/2013MNRAS.428.1055A}
}

@ARTICLE{Sales2012,
       author = {{Sales}, Laura V. and {Navarro}, Julio F. and {Theuns}, Tom and {Schaye}, Joop and {White}, Simon D.~M. and {Frenk}, Carlos S. and {Crain}, Robert A. and {Dalla Vecchia}, Claudio},
        title = "{The origin of discs and spheroids in simulated galaxies}",
      journal = {\mnras},
         year = 2012,
        month = jun,
       volume = {423},
       number = {2},
        pages = {1544-1555},
          doi = {10.1111/j.1365-2966.2012.20975.x},
archivePrefix = {arXiv},
       eprint = {1112.2220},
 primaryClass = {astro-ph.CO},
       adsurl = {https://ui.adsabs.harvard.edu/abs/2012MNRAS.423.1544S}
}

@ARTICLE{White1991,
       author = {{White}, Simon D.~M. and {Frenk}, Carlos S.},
        title = "{Galaxy Formation through Hierarchical Clustering}",
      journal = {\apj},
         year = 1991,
        month = sep,
       volume = {379},
        pages = {52},
          doi = {10.1086/170483},
       adsurl = {https://ui.adsabs.harvard.edu/abs/1991ApJ...379...52W}
}
\bibliographystyle{mnras}

\appendix
\section{Additional Figures}
\label{app:all-sims}

\begin{figure*}
    \centering
    \includegraphics[width=\textwidth]{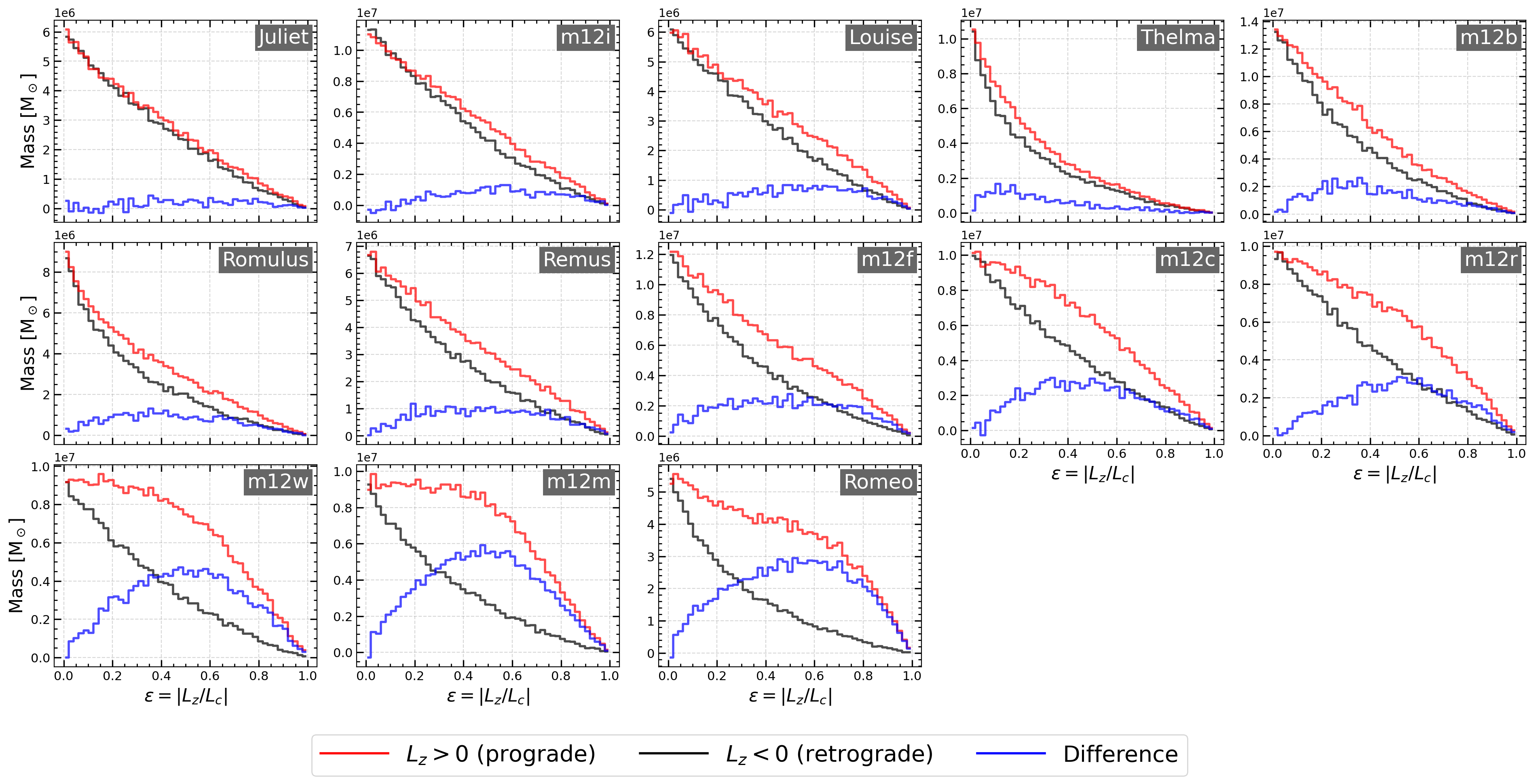}
    \caption{
        Same as Fig.~\ref{fig:flip}, but showing all 13 FIRE-2 simulations. Sorted by increasing diskiness fraction.
    }
    \label{fig:app_eps_all_sims}
\end{figure*}

\begin{figure*}
    \centering
    \includegraphics[width=\textwidth]{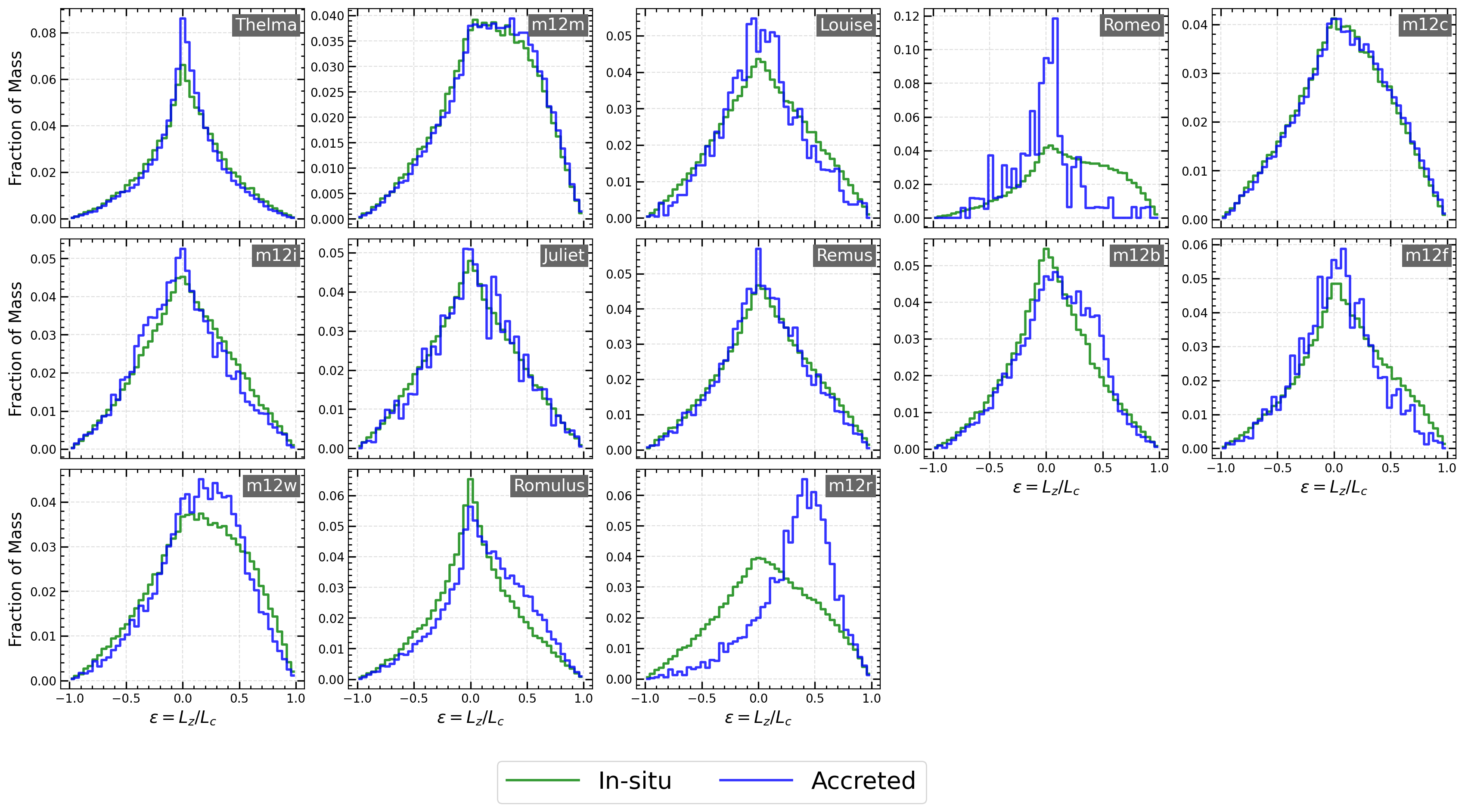}
    \caption{
        Same as Fig.~\ref{fig:massdisk}, but showing all 13 FIRE-2 simulations. Sorted by increasing central stellar surface density ($\Sigma_{\mathrm{cen}}$).
    }
    \label{fig:eps_histos_all}
\end{figure*}

\bsp	
\label{lastpage}
\end{document}